# HIV-1 protease cleavage sites detection with a Quantum convolutional neural network algorithm

Junggu Choi[1,2*], Junho Lee[1,2†], Kyle L. Jung[1,2†], Jae U. Jung[1,2*]
{choij14, leej101, jungk3, jungj}@ccf.org

[1]Department of Microbial Sciences in Health and [2]Global Center for Pathogen and Human Health Research, Cleveland Clinic Research, Cleveland Clinic, OH 44915, USA

[*]Corresponding author: Junggu Choi (choij@ccf.org), Jae U. Jung (jungj@ccf.org)
[†]These authors contributed equally to this work and are listed in alphabetical order.

In this study, we propose a quantum convolutional neural network (QCNN)-based framework with the neural quantum embedding (NQE) to predict HIV-1 protease cleavage sites in amino acid sequences from viral and human proteins. To assess the effectiveness and robustness of our framework, we compared the classification performance against classical neural networks under both noiseless and noisy simulations. Among experimental conditions, the QCNN with the angle and amplitude encoding NQE conditions consistently outperformed classical counterparts in both the similar trainable parameter scale and the different number of qubits (the averaged performance of the 4-qubits and 8-qubits QCNN: 0.9146 and 0.8929 / the averaged performance of the classical neural network: 0.6125 and 0.8278). The QCNN with the NQE showed stable performance under the quantum hardware noise, confirming its applicability to biomedical data analysis with the noise intermediate-scale quantum (NISQ) hardware. This study presents the first application of NQE-augmented QCNNs for HIV-1 cleavage site classification, providing new insights into scalable and noise-resilient quantum machine learning for biomedical data.

## 1. Introduction

Despite decades of global efforts, human immunodeficiency virus (HIV) remains one of the most persistent and adaptive viral threats to public health. According to the World Health Organization (WHO) report published in July 2024, an estimated 39.9 million people are currently living with HIV, and approximately 630,000 deaths were attributed to HIV-related illnesses [1]. Beyond the immediate impact of infection, the progression to advanced HIV disease which is characterized by severe immunodeficiency has been identified as a critical stage where preventive or therapeutic interventions may be most effective. Previous studies have emphasized the importance of early detection and targeted screening strategies to mitigate the risk of severe illness and mortality [2]. In particular, for acquired immunodeficiency syndrome (AIDS), one of the most well-known advanced HIV diseases, a wide range of preventive approaches, including antiretroviral therapies, have been proposed and implemented [3–5].

Among various strategies for the prevention of HIV, analyzing the molecular pathways involved in viral infection has received considerable attention as a basis for therapeutic development. Especially, the biochemical mechanisms of human immunodeficiency virus type 1 (HIV-1) have been extensively studied to identify the key steps involved in viral entry and replication. For example, HIV-1 is known to preferentially infect CD4-positive T cells by binding to the CD4 receptor on the cell surface [6]. Among the molecular processes associated with HIV-1 infection, HIV-1 protease has been identified as an essential enzyme for viral maturation and replication [7, 8]. This enzyme facilitates the cleavage of both viral and host precursor proteins, enabling the formation of mature and functional viral proteins required for infectivity [9]. Due to its essential role, HIV-1 protease and its cleavage sites have been widely investigated as potential therapeutic targets, particularly for the development of protease inhibitors [10–12].

To identify the cleavage sites targeted by HIV-1 protease, researchers initially employed biochemical approaches. For example, four purified human complement proteins (C1q, C2, C4, and C1-inhibitor) involved in the classical complement pathway were incubated *in vitro* with recombinant HIV-1 protease [13]. The cleavage products were analyzed using sodium dodecyl sulfate–polyacrylamide gel electrophoresis (SDS-PAGE) and immunoblotting assay. Sequence analysis revealed that only the N-terminal region of C1-inhibitor contained a protease cleavage site. To further investigate the substrate specificity of HIV-1 protease, researchers applied amino acid replacements at key cleavage regions within the group-specific antigen (Gag) polyprotein, specifically at the p17–p24 and p24–p15 junctions [14]. The reduced cleavage efficiency observed in these modified substrates was confirmed through SDS-PAGE and fluorography. These findings indicate that HIV-1 protease recognizes complex, sequence-dependent characteristics in its cleavage process.

Motivated by previous biological findings and recent progress in machine learning (ML), various ML-based approaches have been proposed to predict HIV-1 protease cleavage sites. To evaluate the effectiveness of different ML algorithms, four models (linear discriminant classifier, LDC; linear support vector machine, LSVM; multilayer perceptron, MLP; and edit distance classifier, EDC;) were applied to the classification task [15]. Based on these models, the authors proposed a hierarchical classification framework that integrates both linear and nonlinear classifiers, achieving a lower error rate (6.6%) compared to the LSVM model alone (9.1%). In a related study, an ensemble learning method incorporating multiple weak learners, including biased SVM classifiers, was also introduced [16]. This ensemble approach demonstrated superior performance, achieving over 92% classification accuracy on three benchmark datasets (1625Dataset, impensDataset, and schillingDataset), outperforming other comparison models.

With the recent advancements in quantum computing technologies, a novel framework known as quantum machine learning (QML) has been proposed as a promising alternative to classical machine learning, aiming to leverage the computational advantages of quantum computing [17, 18]. QML has demonstrated improvements in both runtime efficiency and predictive performance over classical ML methods in various domains [19–21]. For instance, in the context of biomedical data analysis, a quantum–classical hybrid model was proposed to classify neural time-course data associated with early mild cognitive impairment [22]. The model achieved competitive classification performance with a significantly reduced number of trainable parameters, highlighting the parameter efficiency of QML models.

Following the aforementioned strengths, QML algorithms have been explored to uncover latent patterns in biomedical datasets. For example, quantum support vector machines (QSVM) and quantum neural networks (QNN) have been applied to genomic data classification tasks [23]. Although these quantum models did not consistently outperform their classical counterparts, they demonstrated the feasibility and potential of QML approaches for biomedical data analysis. Additionally, QSVM with a ZZ feature map kernel was employed for ligand-based virtual drug screening [24]. Two compound datasets (LIT-PCBA and COVID-19 dataset) were analyzed, containing small molecules labeled as activators or inactivators. Under certain conditions, QSVM outperformed classical SVM in classification accuracy. Furthermore, the quantum algorithm was tested on noisy IBM quantum hardware, and even in the presence of hardware-induced noise, it showed improved performance over classical baselines. These results suggest that QML holds promise for complex biomedical tasks, including drug discovery and molecular classification.

Building on recent examples of QML applications in biomedical data analysis, this study applied a quantum convolutional neural network (QCNN) model to the task of HIV-1 protease cleavage site classification [25]. The research hypothesis was that the QCNN algorithm outperforms a classical neural network model with a similar number of trainable parameters in this classification task. To evaluate this hypothesis, we tested the performance of QCNN under 55 experimental conditions, varying the quantum ansatz and the number of qubits (4 and 8 qubits). In addition, to examine the effect of quantum embedding strategies on model performance, we compared three conventional embedding methods (amplitude encoding, angle encoding, and ZZ feature map) with a recently proposed classical neural network-based approach known as neural quantum embedding (NQE) [26]. Furthermore, the models were evaluated in both noiseless and noisy simulation (IBM FakeBrisbane backend) to assess the robustness of performance under the noisy environment. The key contributions of this study are as follows:

- We systematically evaluated the QCNN algorithm for HIV-1 protease cleavage site classification across 55 experimental conditions.
- To assess the robustness and consistency of quantum algorithm performance, QCNN models were tested in both noiseless and noisy simulations using the fake IBM backend.
- In noiseless simulations, QCNN demonstrated improved performance with the amplitude and angle encoding NQE compared to classical counterparts.
- Quantum algorithms including NQE and QCNN showed higher performance than classical counterparts under the noisy environment including quantum hardware noises.

The remainder of this paper is structured as follows. Section 2 presents the research framework, including descriptions of the four HIV-1 protease cleavage site datasets (746, 1625, Schilling, and Impens), the quantum algorithms (NQE and QCNN), and the evaluation metrics used for classification performance. Section 3 outlines the experimental results, detailing the classification performance under each experimental condition in both noiseless simulation and noisy quantum hardware settings. Section 4 discusses the potential of QCNN for HIV-1 protease cleavage site detection, as well as observed trends across the experimental conditions. Finally, Section 5 concludes the paper by summarizing key findings and highlighting the strengths and limitations of this study.

## 2. Methods

The overall research design of this study is summarized in Figure 1, comprising three sequential stages. First, four publicly available HIV-1 protease cleavage site datasets (746, 1625, Schilling, and Impens dataset) were collected. Second, in the preprocessing stage, the four datasets were merged into a single dataset, and duplicated sequences were removed. The resulting sequences were then transformed into numerical features using one-hot encoding of amino acid sequences. Finally, the classification performance of the QCNN was evaluated under various experimental conditions and compared with a classical neural network.

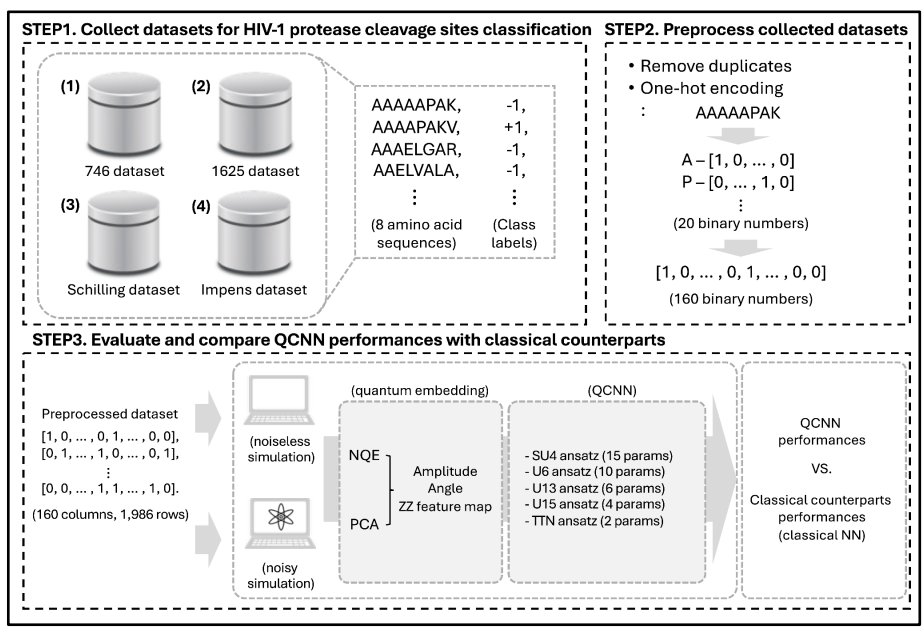

Figure 1. The research scheme of this study

## 2.1 Datasets

To evaluate the performance of the QCNN algorithm and its classical counterparts, four open-source HIV-1 protease cleavage site datasets were collected. The first dataset, referred to as the 746 dataset, was published in a previous study that employed SVM and multilayer perceptrons (MLP) to analyze the specificity of HIV-1 protease [27]. It contains 746 sequences that were experimentally validated in earlier studies. All sequences were derived from viral proteins, including Gag and Pol protein. The second dataset, known as the 1625 dataset, comprises 1,625 sequences collected from prior literature sources [28]. This dataset was compiled for computational analysis of the HIV-1 protease interactome, and most of the sequences originate from viral proteins.

In contrast to the 746 and 1625 datasets which consist of sequences from viral proteins, the Schilling dataset includes 3,272 cleavage site sequences derived from host proteins (i.e., human) [29]. To identify these sequences, the authors utilized the proteomic identification of cleavage sites (PICS) method applied to an oligopeptide library. These cleavage sites were identified based on exploratory analyses, rather than through strict *in vivo* biological validation. Lastly, the Impens dataset contains 947 cleavage site sequences obtained through a proteomics-based investigation [30]. For this dataset, the authors assessed the potential of HIV-1 protease to recognize host protein cleavage sites using N-terminomics-based profiling. The sequences were identified through in vitro protease treatment, followed by mass spectrometry analysis.

## 2.2 Preprocessing

From the four collected datasets, all sequences consisting of eight amino acids were merged into a single dataset to maximize sample utilization. The initial merged dataset comprised a total of 6,590 samples (746 from the 746 dataset, 1,625 from the 1625 dataset, 3,272 from the Schilling dataset, and 947 from the Impens dataset). Duplicate entries were subsequently removed, reducing the dataset size to 5,840 unique samples. At this stage, the distribution between the two classes was imbalanced, with 4,847 non-cleavage site sequences (class label: –1) and 993 cleavage site sequences (class label: +1). To address this class imbalance, 993 non-cleavage site samples were randomly undersampled to match the number of cleavage site samples. Following class balancing, one-hot encoding was applied to convert the 8 amino acid sequences into numerical format suitable for machine learning evaluation. Each amino acid was represented as a binary vector of length 20, corresponding to the 20 standard amino acids. For instance, the character 'A' (Alanine) was encoded as [1, 0, ..., 0]. As a result, each sequence was transformed into a binary vector of length 160, and the final dataset for model evaluation consisted of 1,986 samples.

## 2.3 Quantum embedding methods

To utilize the quantum algorithm for evaluating the preprocessed classical binary vectors, it is necessary to encode the classical data into quantum states. In this study, we compared four quantum embedding methods to investigate how the choice of embedding influences algorithm performance. As the first approach, we adopted the amplitude encoding technique, a widely used quantum embedding method. This method encodes the $N$-dimensional classical input vector $x = (x_1, x_2, \ldots, x_{N-1}, x_N)^T$ into the amplitude components of an $n$-qubits quantum state. The amplitude encoding function is defined as follows:

$$U_\phi(x): x \in \mathbb{R}^N \to |\phi(x)\rangle = \frac{1}{\|x\|} \sum_{i=1}^{N} x_i |i\rangle \quad (1)$$

In Equation (1), $|\phi(x)\rangle$ denotes the resulting quantum state and $|i\rangle$ represents the computational basis state.

As the second embedding method, the angle encoding method was employed. This method encodes each classical feature value to the phase of a single-qubit quantum state by rescaling the input from 0 to $\pi$. The encoding function for angle encoding is defined as follows:

$$U_\phi(x): x \in \mathbb{R}^N \to |\phi(x)\rangle = \otimes_{i=1}^{N} (\cos\left(\frac{x_i}{2}\right)|0\rangle + \sin\left(\frac{x_i}{2}\right)|1\rangle) \quad (2)$$

Third, the ZZ feature map was utilized, which incorporates linear entanglement through controlled rotation gates [32]. In this study, the linear ZZ feature map was selected among the three possible configurations (linear, circular, and fully connected) to evaluate a simpler structure. The circuit design for the 4-qubit linear ZZ feature map is illustrated in Figure 2.

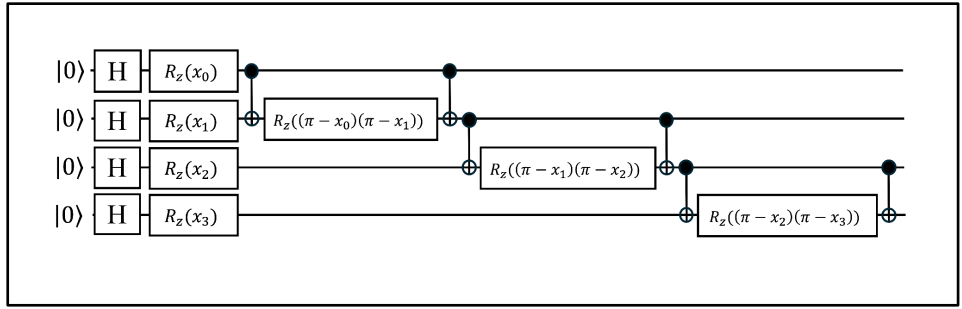

Figure 2. The circuit structure of the 4-qubits linear ZZ feature map

Finally, the Neural Quantum Embedding (NQE) method, which incorporates a classical neural network within the quantum embedding framework, was used. Analogous to the concept of kernel-target alignment in QML, this method optimizes the quantum embedding by training a classical neural network. The NQE framework comprises two sequential modules: the first is a classical neural network that processes the input features, and the second is a quantum embedding module such as the ZZ feature map, amplitude encoding, or angle encoding. The objective of NQE is to minimize the mean squared error between the quantum fidelity, computed from two quantum states encoded from a pair of classical input features, and the modified class label derived from the corresponding two original class labels. The fidelity-based loss function used in NQE is defined as follows:

$$L = \underset{w}{\mathrm{argmin}} \sum_{i,j} \left[ |\langle x_i(w)|x_j(w)\rangle|^2 - \frac{1}{2}(1 + y_i y_j) \right]^2 \quad (3)$$

In Equation (3), $w$ represents the weight vector of the classical neural network in the NQE framework, and $x_i$, $y_i$ denote the data sample and its corresponding class label, respectively. As an example of the optimization procedure in NQE, when two classical input features are sampled from the same class (i.e., $y_i = -1$, $y_j = -1$ or $y_i = +1$, $y_j = +1$), the target value of the second term in the mean squared error becomes 1. Ideally, the quantum fidelity between quantum states encoded from the same class is close to 1. Therefore, the NQE is trained to produce fidelity values near 1 for samples from the same class and near 0 for samples from different classes. Upon completion of training, the effectiveness of NQE can be evaluated by comparing the trace distance ($\frac{1}{2}\|\rho - \sigma\|_1$), which quantifies the distinguishability between two quantum states, before and after training.

## 2.4 Quantum convolutional neural network algorithm

In this study, the QCNN algorithm was employed as the quantum model. Initially introduced by previous work [25], QCNN was designed to operate effectively on Noisy Intermediate-Scale Quantum (NISQ) devices, which represent the most practically accessible quantum hardware to date [32]. Due to inherent noise and hardware constraints in NISQ systems, hybrid quantum–classical models using compact quantum circuits have become a practical direction in quantum computing research. QCNN was selected for this study based on several advantages. First, its structure as a variational quantum algorithm (VQA) is well-suited for NISQ implementation, as it consists of shallow circuit depths and relies solely on nearest-neighbor two-qubit gates, aligning with the limited qubit connectivity of current hardware. Second, the QCNN architecture reduces the number of qubits as the circuit deepens and shares parameters across layers, features that help mitigate the barren plateau problem, an issue in quantum machine learning where gradients vanish during training, making optimization difficult [33]. Third, previous studies have demonstrated the effectiveness of QCNN across a variety of tasks in both classical and quantum domains [34, 35]. The overall architecture of the

QCNN model is illustrated in Figure 3. The purple section represents the NQE module, which encodes classical input data into quantum states using a classical neural network. The gray section corresponds to the structure of the QCNN.

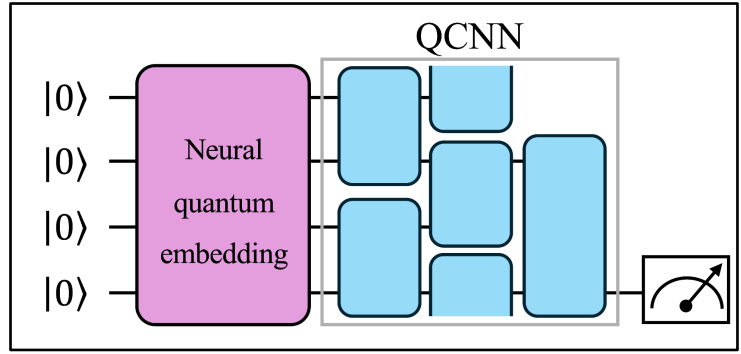

Figure 3. The scheme of the 4-qubits QCNN model with NQE

To evaluate the influence of ansatz selection on algorithm performance, we adopted five different ansatz types (TTN, U15, U13, U6, and SU4 gates), all of which have been used in previous studies for constructing quantum convolutional layers. Since each ansatz involves a different number of trainable parameters, model performance was assessed in relation to the scale of the trainable parameter. The specific configurations and circuit structures of these five ansatzes are summarized in Table 1 and illustrated in Figure 4. To further explore the performance of simplified QCNN architectures, we also implemented QCNN without a quantum pooling layer. Additionally, while previous studies used mean squared error (MSE) and cross-entropy (CE) loss functions for training QCNN models [25], [26] introduced a linear loss function for training QCNN models integrated with pretrained NQE. The linear loss function is defined as follows:

$$L = \frac{1}{n}\sum_{i=1}^{N}\frac{1}{2}(1 - y_i \hat{y}_i) \tag{4}$$

Here, $y_i \in \{-1, +1\}$ denotes the true label of the $i$-th sample, $\hat{y}_i \in [-1, +1]$ represents the predicted value for that sample, and $N$ is the total number of samples. Based on the Helstrom bound, the linear loss function can be bounded below by an inequality involving the trace distance, specifically $linear\ loss\ function \geq \frac{1}{2}(1 - D_{tr})$. This inequality indicates that the classification error calculated from the linear loss function can be bounded by the trace distance. From this inequality, we applied the linear loss function to train the QCNN in order to evaluate how changes in trace distance relate to model performance.

Table 1. Description of five ansatz for the quantum convolutional layer

| No. | Ansatz | The number of parameters | Reference |
| --- | --- | --- | --- |
| 1 | SU4 gate | 15 trainable parameters | [36, 37] |
| 2 | U6 gate | 10 trainable parameters | [38] |
| 3 | U13 gate | 6 trainable parameters | [38] |
| 4 | U15 gate | 4 trainable parameters | [38] |
| 5 | TTN gate | 2 trainable parameters | [39] |

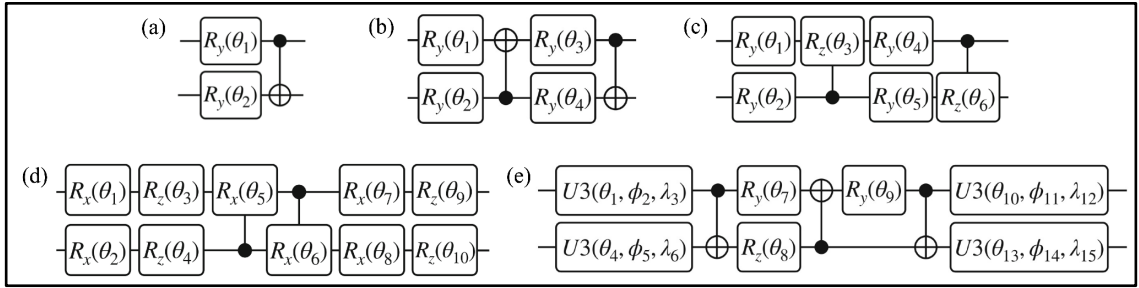

Figure 4. Circuit structures of five ansatz for the quantum convolutional layer ((a): TTN gate, (b): U15 gate, (c): U13 gate, (d): U6 gate, (e): SU4 gate)

## 2.5 Classical counterpart algorithms

To assess the appropriateness of using the QCNN algorithm for this classification task, its performance was compared with that of classical models. As a classical counterpart of QCNN, a classical neural network with a comparable number of trainable parameters was implemented to evaluate the computational efficiency of the quantum model. This network was trained using the binary cross entropy loss function, as the task involves binary classification. Additionally, to provide a fair comparison with the quantum model using NQE, we trained a classical neural network using the fidelity-based loss function without the quantum part. The trained network was then used as a feature extractor for a second classical neural network configured with a similar number of trainable parameters to those used in the QCNN.

## 2.6 Model evaluations

To evaluate the performance of the QCNN model in classifying HIV-1 protease cleavage sites, a total of 55 experimental conditions were designed. These conditions were defined by varying the number of qubits (4 and 8), the number of trainable parameters associated with five different ansatz structures, and the type of quantum embedding method, either NQE or one of the conventional techniques. For experimental conditions without NQE, principal component analysis (PCA) was applied to reduce the dimensionality of the one-hot encoded input features to match the QCNN input dimension. A detailed overview of the experimental configurations is provided in Table 2 (for NQE-based conditions) and Table 3 (for PCA-based conditions, i.e., without NQE).

Table 2. Experimental conditions for the QCNN model with NQE

| No. | # of qubits | Embedding methods | Ansatz for QCNN[1] (# of total trainable parameters) | No. | # of qubits | Embedding methods | Ansatz for QCNN (# of total trainable parameters) |
|---|---|---|---|---|---|---|---|
| 1 | 4 | NQE+ZZ[2] | SU4 gate (30 trainable parameters) | 16 | 8 | NQE+ZZ | SU4 gate (45 trainable parameters) |
| 2 | | | U6 gate (20 trainable parameters) | 17 | | | U6 gate (30 trainable parameters) |
| 3 | | | U13 gate (12 trainable parameters) | 18 | | | U13 gate (18 trainable parameters) |
| 4 | | | U15 gate (8 trainable parameters) | 19 | | | U15 gate (12 trainable parameters) |
| 5 | | | TTN gate (4 trainable parameters) | 20 | | | TTN gate (6 trainable parameters) |
| 6 | | NQE+Amp[3] | SU4 gate (30 trainable parameters) | 21 | | NQE+Amp | SU4 gate (45 trainable parameters) |
| 7 | | | U6 gate (20 trainable parameters) | 22 | | | U6 gate (30 trainable parameters) |
| 8 | | | U13 gate (12 trainable parameters) | 23 | | | U13 gate (18 trainable parameters) |
| 9 | | | U15 gate (8 trainable parameters) | 24 | | | U15 gate (12 trainable parameters) |
| 10 | | | TTN gate (4 trainable parameters) | 25 | | | TTN gate (6 trainable parameters) |
| 11 | | NQE+Ang[4] | SU4 gate (30 trainable parameters) | 26 | | NQE+Ang | SU4 gate (45 trainable parameters) |
| 12 | | | U6 gate (20 trainable parameters) | 27 | | | U6 gate (30 trainable parameters) |
| 13 | | | U13 gate (12 trainable parameters) | 28 | | | U13 gate (18 trainable parameters) |
| 14 | | | U15 gate (8 trainable parameters) | 29 | | | U15 gate (12 trainable parameters) |
| 15 | | | TTN gate (4 trainable parameters) | 30 | | | TTN gate (6 trainable parameters) |

[1]QCNN: the quantum convolutional neural network; [2]NQE+ZZ: the ZZ feature map with the neural quantum embedding method; [3]NQE+Amp: the amplitude encoding with the neural quantum embedding method; [4]NQE+Ang: the angle encoding with the neural quantum embedding method;

Table 3. Experimental conditions for the QCNN model with PCA (without NQE)

| No. | # of qubits | Embedding methods | Ansatz for QCNN[1] (# of total trainable parameters) | No. | # of qubits | Embedding methods | Ansatz for QCNN (# of total trainable parameters) |
|---|---|---|---|---|---|---|---|
| 1 | 4 | PCA+ZZ[2] | SU4 gate (30 trainable parameters) | 16 | 8 | PCA+ZZ | SU4 gate (45 trainable parameters) |
| 2 | | | U6 gate (20 trainable parameters) | 17 | | | U6 gate (30 trainable parameters) |
| 3 | | | U13 gate (12 trainable parameters) | 18 | | | U13 gate (18 trainable parameters) |
| 4 | | | U15 gate (8 trainable parameters) | 19 | | | U15 gate (12 trainable parameters) |
| 5 | | | TTN gate (4 trainable parameters) | 20 | | | TTN gate (6 trainable parameters) |
| 6 | | PCA+Amp[3] | SU4 gate (30 trainable parameters) | 21 | | NQE+Ang | SU4 gate (45 trainable parameters) |
| 7 | | | U6 gate (20 trainable parameters) | 22 | | | U6 gate (30 trainable parameters) |
| 8 | | | U13 gate (12 trainable parameters) | 23 | | | U13 gate (18 trainable parameters) |
| 9 | | | U15 gate (8 trainable parameters) | 24 | | | U15 gate (12 trainable parameters) |
| 10 | | | TTN gate (4 trainable parameters) | 25 | | | TTN gate (6 trainable parameters) |
| 11 | | PCA+Ang[4] | SU4 gate (30 trainable parameters) | | | | |
| 12 | | | U6 gate (20 trainable parameters) | | | | |
| 13 | | | U13 gate (12 trainable parameters) | | | | |
| 14 | | | U15 gate (8 trainable parameters) | | | | |
| 15 | | | TTN gate (4 trainable parameters) | | | | |

[1]QCNN, the quantum convolutional neural network; [2]PCA+ZZ, the ZZ feature map with the principal components analysis; [3]PCA+Amp, the amplitude encoding with the principal components analysis; [4]PCA+Ang, the angle encoding with the principal components analysis;

For training both the NQE and QCNN models, a mini batch of samples was randomly selected from the training dataset at each iteration. This mini-batch approach not only reduced the simulation time compared to full-batch training but also improved optimization by enabling gradients to escape local minima. In the case of NQE, an early stopping strategy was employed, terminating the training process if validation loss value failed to improve for 40 consecutive iterations. In this process, the validation value is calculated per 10 iterations, not all iterations. The Adam optimizer was used for training the NQE model, while Nesterov momentum was applied to optimize the QCNN. In addition, three hyperparameters (the number of training iterations, learning rate, and batch size) were optimized under the random selection. Furthermore, to consider the possible variations of the model performance from the mini-batch sampling, we repeated five times and compared mean and standard deviation values for all experimental results.

## 2.8 Tools

Data preprocessing and result visualization were performed using Python (version 3.10), ensuring a uniform computational environment throughout the entire analysis. All models, including the NQE, QCNN, classical neural network, and classical SVM were implemented using PennyLane (version 0.36.0), PyTorch (version 2.5.1), and Scikit-learn (version 1.7.1). The experiments for the noisy simulation were implemented using Qiskit (version 1.2.4).

# 3. Results

## 3.1 Training results of NQE in the noiseless simulation

To maximize the trace distance between two quantum states which encoding amino acid sequence information, we trained NQE models first with the hyperparameter tuning. The structure of the classical neural network in NQE was optimized with the fixed input and output layer length in their structure to consider the input vector length and the number of classical input features for the following quantum embedding method. The optimized classical neural network structures and the hyperparameters such as the batch size, learning rate, and the number of iterations are listed in Table 5. Following the optimized hyperparameters and model structures, we checked the increased trace distance value through the comparison between before NQE training and after NQE training. Among six NQE conditions we trained, five NQE conditions except NQE with the 4-qubits ZZ feature map showed the improvement of the trace distance value close to 1. Furthermore, when we checked the training and validation loss graph after NQE training to check the stable convergence of the NQE model, we found that training and validation loss graphs of all NQE conditions converged stably. The detailed trace distance values and loss graphs are listed in Figure 5, 6, and Table 4.

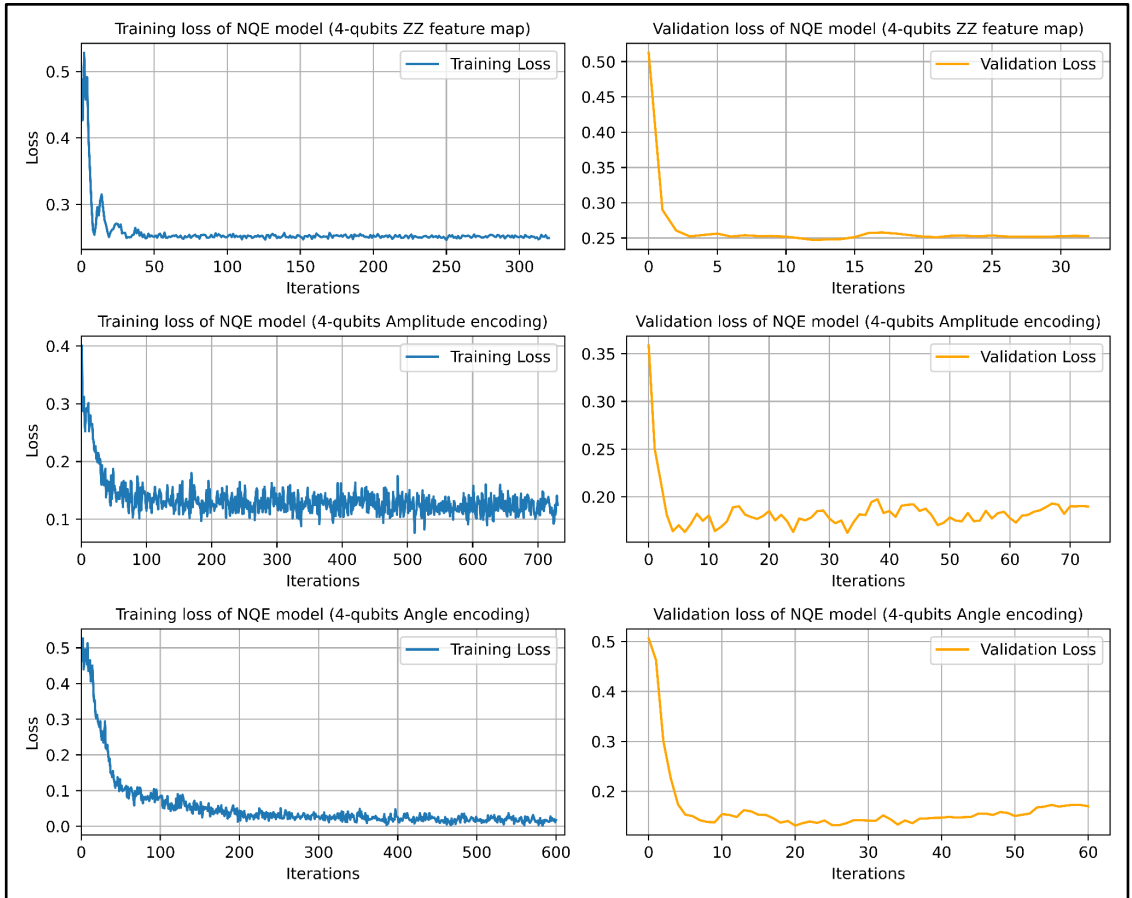

Figure 5. Training and validation loss graphs of 4-qubits NQE in noiseless simulations

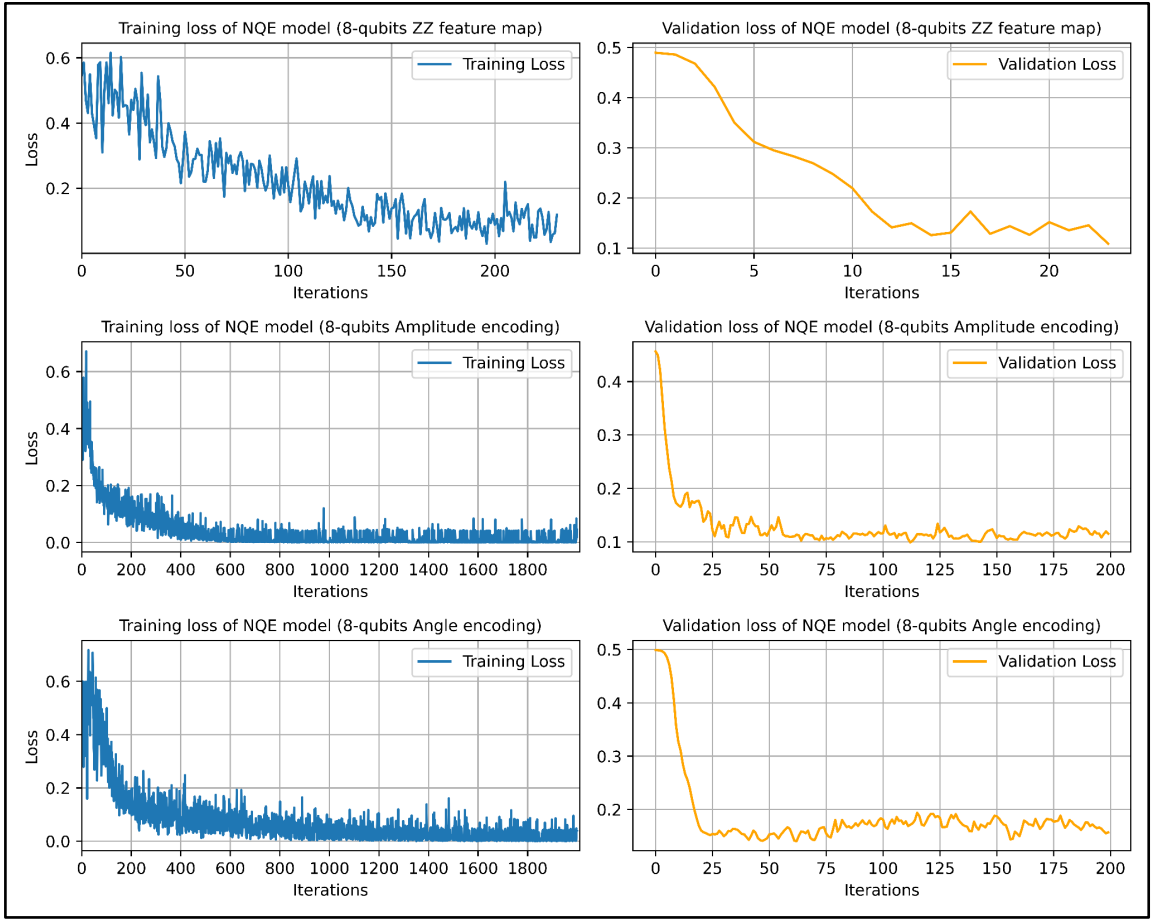

Figure 6. Training and validation loss graphs of 8-qubits NQE in noiseless simulations

Table 4. Trace distance values calculated from NQE for the training and test set (before vs. after NQE training) in noiseless simulations

| No. | NQE condition | Trace distance of training set (before NQE training) | Trace distance of test set (before NQE training) | Trace distance of training set (after NQE training) | Trace distance of test set (after NQE training) |
|---|---|---|---|---|---|
| 1 | 4-qubits ZZ[1] | 0.0641 | 0.0713 | 0.0854 | 0.0986 |
| 2 | 4-qubits Amp[2] | 0.0253 | 0.0314 | 0.9972 | 0.7642 |
| 3 | 4-qubits Ang[3] | 0.0010 | 0.0003 | 0.9695 | 0.7671 |
| 4 | 8-qubits ZZ | 0.1509 | 0.1855 | 0.8154 | 0.7296 |
| 5 | 8-qubits Amp | 0.1388 | 0.1768 | 0.9634 | 0.7569 |
| 6 | 8-qubits Ang | 0.0006 | 0.0017 | 0.9820 | 0.7923 |

[1]ZZ: the ZZ feature map; [2]Amp: the amplitude encoding; [3]Ang: the angle encoding;

Table 5. Classical neural network structures for NQE and hyperparameters for NQE training in the noise simulation

| No. | NQE condition | Classical neural network structure for NQE | Hyperparameters for NQE training |
|---|---|---|---|
| 1 | 4-qubits ZZ feature map | nn.Linear(160, 8),<br>nn.ReLU(),<br>nn.Linear(8, 8) | - Batch size: 512<br>- Iterations: 2000<br>- learning rate: 0.01 |
| 2 | 4-qubits amplitude encoding | nn.Linear(160, 64),<br>nn.ReLU(),<br>nn.Linear(64, 32),<br>nn.ReLU(),<br>nn.Linear(32, 16) | - Batch size: 256<br>- iterations: 2000<br>- learning rate: 0.01 |
| 3 | 4-qubits angle encoding | nn.Linear(160, 16),<br>nn.ReLU(),<br>nn.Linear(16, 8),<br>nn.ReLU(),<br>nn.Linear(8, 4) | - Batch size: 256<br>- iterations: 2000<br>- learning rate: 0.008 |
| 4 | 8-qubits ZZ feature map | nn.Linear(160, 64),<br>nn.ReLU(),<br>nn.Linear(64, 32),<br>nn.ReLU(),<br>nn.Linear(32, 16) | - Batch size: 25<br>- iterations: 2000<br>- learning rate: 0.01 |
| 5 | 8-qubits amplitude encoding | nn.Linear(160, 80),<br>nn.ReLU(),<br>nn.Linear(80, 256) | - Batch size: 25<br>- iterations: 2000<br>- learning rate: 0.02 |
| 6 | 8-qubits angle encoding | nn.Linear(160, 16),<br>nn.ReLU(),<br>nn.Linear(16, 8) | - Batch size: 25<br>- iterations: 2000<br>- learning rate: 0.01 |

### 3.2 Training results of NQE in the noisy simulation

To assess the effect of hardware noise on the training of the NQE model, we conducted additional evaluations using a noisy simulation environment based on the IBM FakeBrisbane backend. This backend replicates the noise characteristics of the 127-qubit IBM quantum processor, IBM Brisbane. While training in a noiseless simulation showed improvements in trace distance, training under noisy conditions resulted in a stabilized trace distance around 0.6 after convergence. The detailed trace distance values obtained from the noisy simulation are summarized in Table 6.

To account for the presence of noise during training, we adjusted the hyperparameters of the NQE, accordingly. The selected hyperparameters for noisy simulations are listed in Table 7. As in the noiseless case, we checked the training and validation loss graphs to ensure the convergence of the NQE models. The corresponding loss graphs for 4-qubit and 8-qubit NQE models under noisy conditions are presented in Figure 7 and Figure 8.

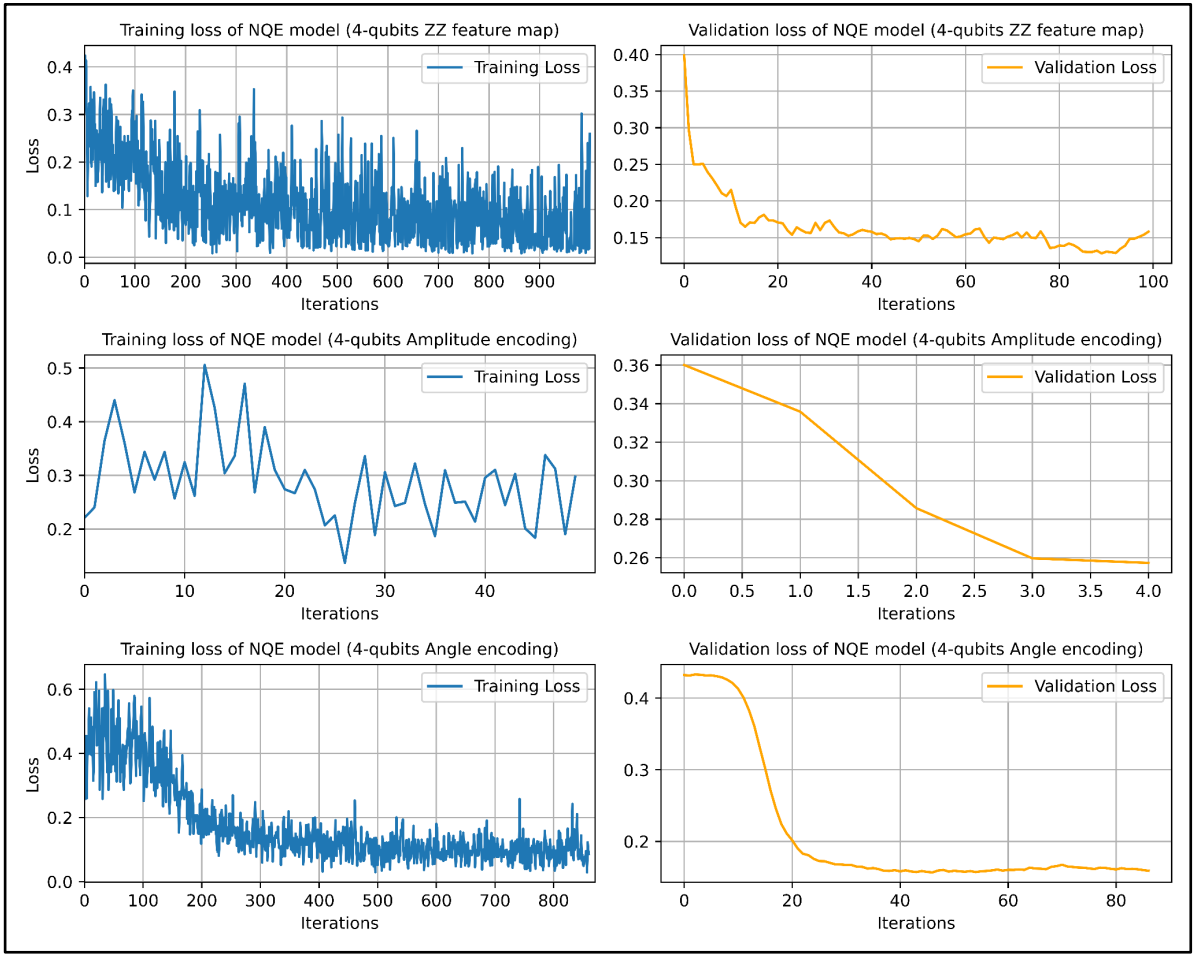

Figure 7. Training and validation loss graphs of 4-qubits NQE in noisy simulations

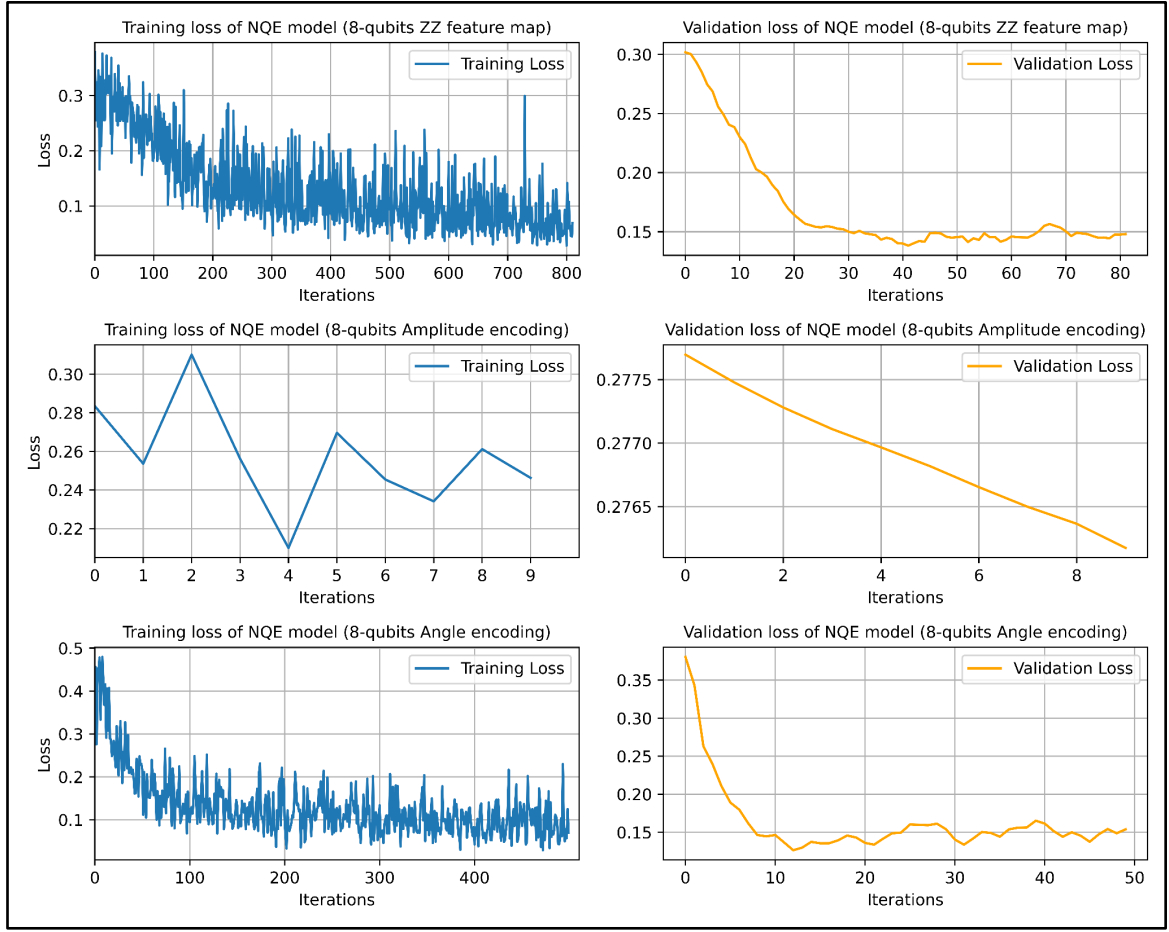

Figure 8. Training and validation loss graphs of 8-qubits NQE in noisy simulations

Table 6. Trace distance values calculated from NQE for the training and test set (before vs. after NQE training) in noisy simulations

| No. | NQE condition | Trace distance of training set (before NQE training) | Trace distance of test set (before NQE training) | Trace distance of training set (after NQE training) | Trace distance of test set (after NQE training) |
|---|---|---|---|---|---|
| 1 | 4-qubits ZZ[1] | 0.0659 | 0.0569 | 0.6658 | 0.5715 |
| 2 | 4-qubits Amp[2] | 0.0280 | 0.0434 | 0.2487 | 0.2339 |
| 3 | 4-qubits Ang[3] | 0.0009 | 2.04e-05 | 0.5963 | 0.5538 |
| 4 | 8-qubits ZZ | 0.1509 | 0.1855 | 0.5362 | 0.4387 |
| 5 | 8-qubits Amp | 0.1621 | 0.1737 | 0.1432 | 0.1858 |
| 6 | 8-qubits Ang | 0.0062 | 0.0041 | 0.6736 | 0.5841 |

[1]ZZ: the ZZ feature map; [2]Amp: the amplitude encoding; [3]Ang: the angle encoding;

Table 7. Classical neural network structures for NQE and hyperparameters for NQE training in the noisy simulation

| No. | NQE condition | Classical neural network structure for NQE | Hyperparameters for NQE training |
|---|---|---|---|
| 1 | 4-qubits ZZ feature map | nn.Linear(160, 8),<br>nn.ReLU(),<br>nn.Linear(8, 8) | - Batch size: 10<br>- Iterations: 1000<br>- learning rate: 0.01 |
| 2 | 4-qubits amplitude encoding | nn.Linear(160, 64),<br>nn.ReLU(),<br>nn.Linear(64, 32),<br>nn.ReLU(),<br>nn.Linear(32, 16) | - Batch size: 10<br>- iterations: 50<br>- learning rate: 0.0008 |
| 3 | 4-qubits angle encoding | nn.Linear(160, 16),<br>nn.ReLU(),<br>nn.Linear(16, 8),<br>nn.ReLU(),<br>nn.Linear(8, 4) | - Batch size: 30<br>- iterations: 1000<br>- learning rate: 0.0005 |
| 4 | 8-qubits ZZ feature map | nn.Linear(160, 64),<br>nn.ReLU(),<br>nn.Linear(64, 32),<br>nn.ReLU(),<br>nn.Linear(32, 16) | - Batch size: 10<br>- iterations: 2000<br>- learning rate: 0.001 |
| 5 | 8-qubits amplitude encoding | nn.Linear(160, 80),<br>nn.ReLU(),<br>nn.Linear(80, 256) | - Batch size: 10<br>- iterations: 10<br>- learning rate: 0.00008 |
| 6 | 8-qubits angle encoding | nn.Linear(160, 16),<br>nn.ReLU(),<br>nn.Linear(16, 8) | - Batch size: 20<br>- iterations: 500<br>- learning rate: 0.01 |

### 3.3 QCNN and classical counterpart performances in the noiseless simulation
*3.3.1 Comparison of Model Performance between QCNN with Pretrained NQE and Classical Counterparts*

To evaluate the applicability of quantum models to the HIV-1 protease cleavage site classification task, we first compared the average classification accuracy of the QCNN model with pretrained NQE to that of classical neural network models, including the classical counterpart of NQE. Among the three 4-qubit NQE configurations (NQE with ZZ feature map, amplitude encoding, and angle encoding) combined with a 4-qubit QCNN, the QCNN models using amplitude encoding (with 20 and 12 trainable parameters) and angle encoding (with 30, 20, and 12 trainable parameters) in NQE yielded higher average accuracy values (0.8649, 0.8844, 0.9246, 0.9226, and 0.9246, respectively) than the classical neural networks. Detailed comparison results for the 4-qubit NQE and QCNN models are shown in Figure 9.

In addition, to assess the performance with a larger number of qubits, we evaluated the 8-qubit quantum models. In this setting, the QCNN models with NQE using the ZZ feature map (with 45 and 18 trainable parameters) showed higher average accuracy values (0.9005 and 0.9030) than the classical neural networks. Similarly, the QCNN models with NQE using angle encoding (with 45, 30, 18, and 12 trainable parameters) also outperformed the classical models, with accuracy values of 0.8940, 0.8945, 0.8915, and 0.8915, respectively (classical neural networks: 0.8493, 0.8494, 0.8489, and 0.7647). The full results of model comparisons for 8-qubit NQE and QCNN are presented in Figure 10. Accuracy values for all 4-qubit and 8-qubit QCNN models, along with classical neural networks, are summarized in Appendix A and B.

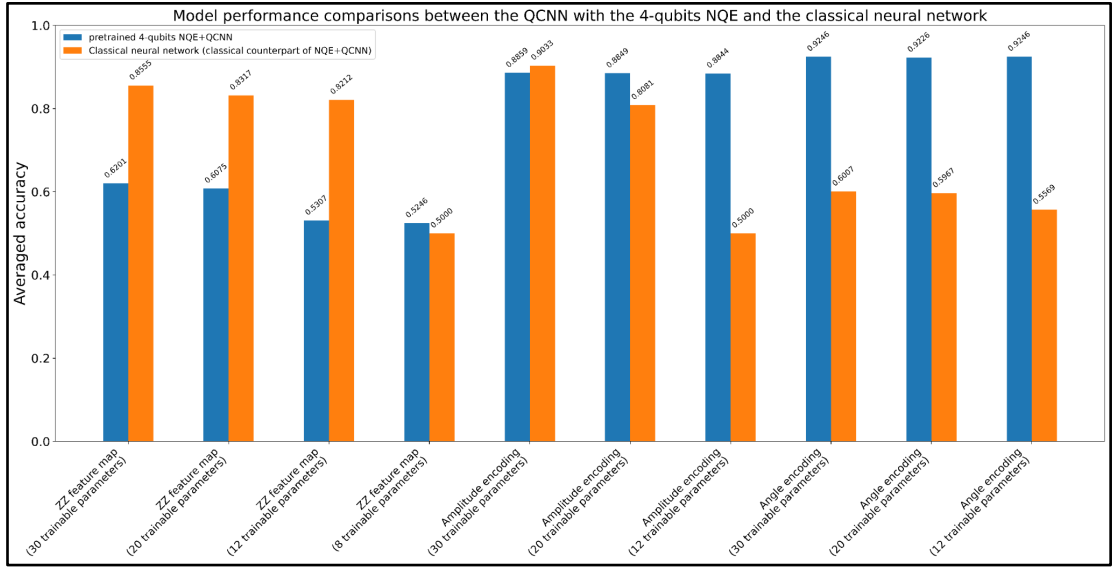

Figure 9. Model performance comparisons between QCNN models with 4-qubits NQE and classical neural network models in the noiseless simulation

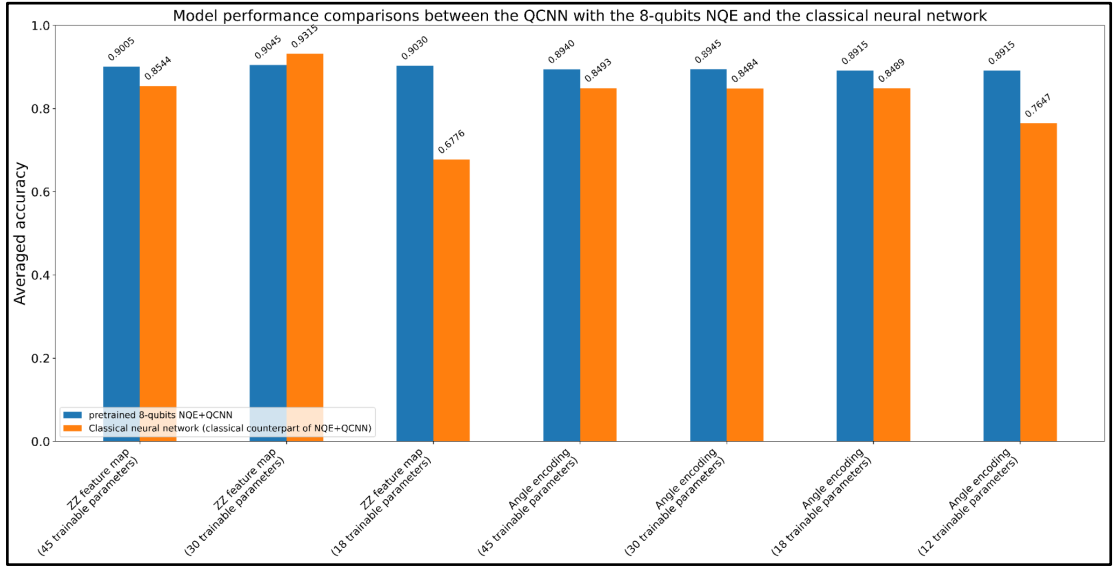

Figure 10. Model performance comparisons between QCNN models with 8-qubits NQE and classical neural network models in the noiseless simulation

*3.3.2 Performance comparison of QCNN models with NQE versus traditional embeddings*

To assess the influence of NQE on model performance, we further evaluated QCNN models without NQE, where PCA was used for dimensionality reduction. In the 4-qubit setting, all QCNN models combined with NQE using amplitude or angle encoding outperformed their counterparts with PCA. However, in the case of the ZZ feature map, only the QCNN model with 8 trainable parameters showed improved accuracy compared to the PCA-based QCNN. For the 8-qubit setting, QCNN models incorporating NQE with either the ZZ feature map or angle encoding consistently outperformed the PCA-based models. The comparison results between NQE and PCA conditions are illustrated in Figure 11 and Figure 12. A complete list of hyperparameter configurations and model performance metrics for 4-qubit and 8-qubit QCNN models with PCA is provided in Appendix C.

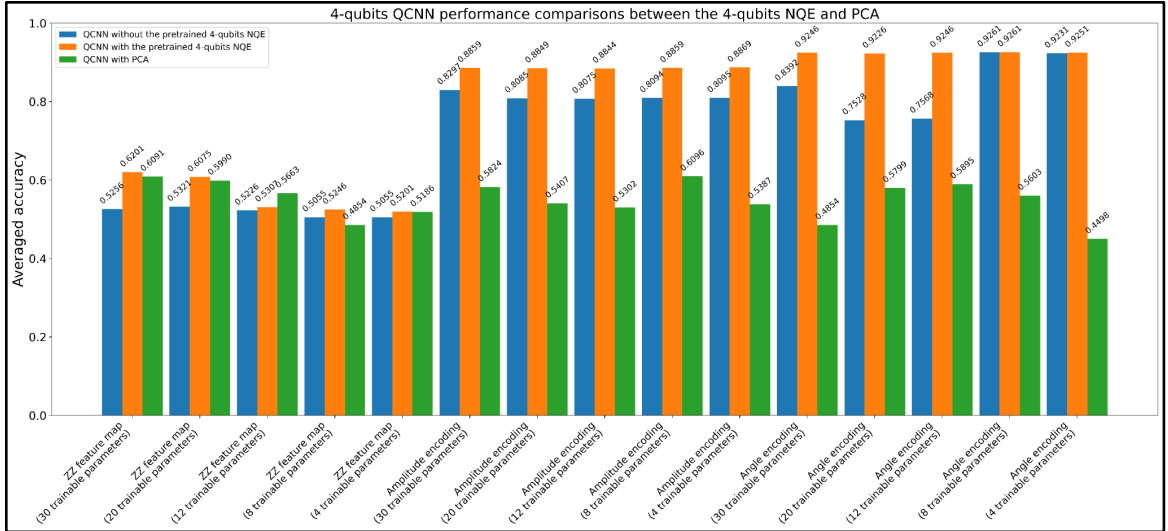

Figure 11. 4-qubits QCNN performances from 4-qubits NQE and PCA conditions in the noiseless simulation

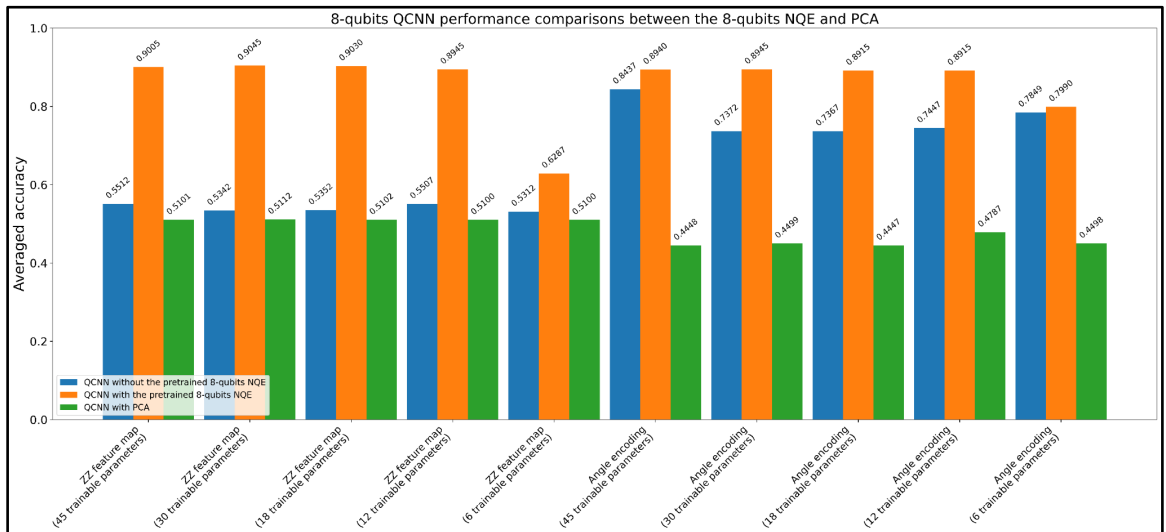

Figure 12. 8-qubits QCNN performances from 8-qubits NQE and PCA conditions in the noiseless simulation

### 3.3.3 Relationship between QCNN performance and the number of trainable parameters

To evaluate how the number of trainable parameters affects model performance, we checked the averaged accuracy values of QCNN models trained with the pretrained NQE. In the 4-qubit QCNN results, models using the ZZ feature map exhibited an increasing trend in accuracy as the number of parameters increased. In contrast, models using amplitude or angle encoding showed relatively stable performance across parameter scales. Similarly, in the 8-qubit QCNN results, the amplitude encoding condition showed no noticeable change in performance. However, for the ZZ feature map and angle encoding, performance improvements were observed when the number of parameters increased from 6 to 12, after which accuracy values stabilized. These trends are illustrated in Figure 13 and Figure 14.

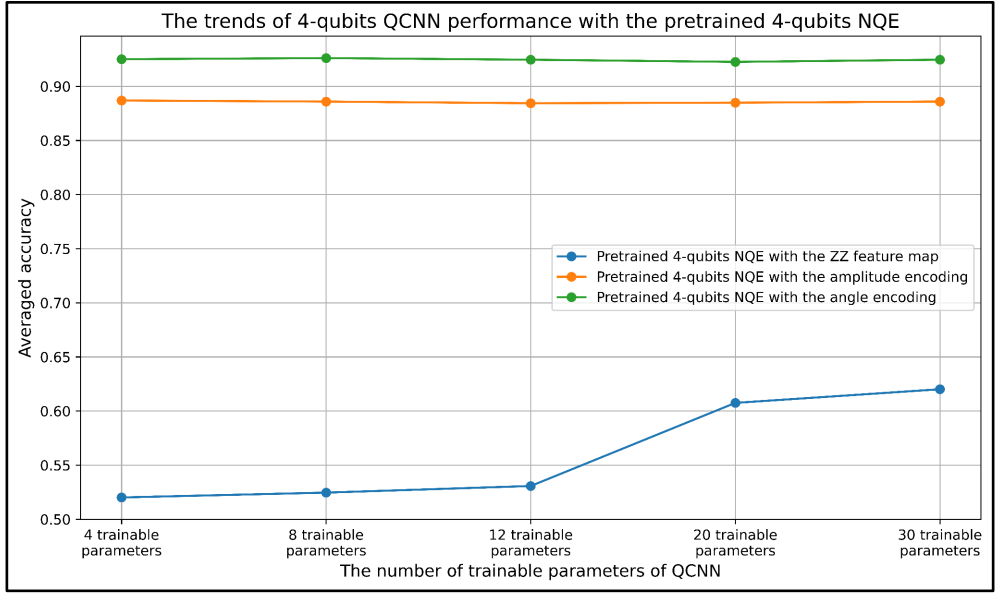

Figure 13. The performance trends of 4-qubits QCNN with the pretrained 4-qubits NQE in the noiseless simulation

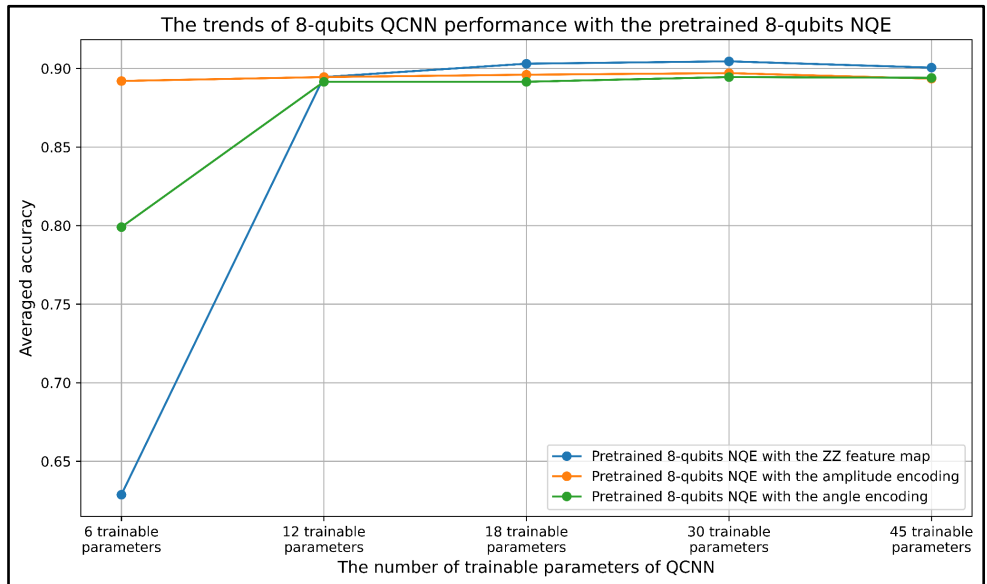

Figure 14. The performance trends of 8-qubits QCNN with the pretrained 8-qubits NQE in the noiseless simulation

### 3.4 QCNN performances in the noisy simulation
#### 3.4.1 Model Performance of QCNN with Pretrained NQE under quantum hardware noises

To examine the effect of quantum hardware noise on the performance of the QCNN model, we extended our evaluation beyond the noiseless simulation by conducting experiments under a noisy environment using the IBM FakeBrisbane backend, which emulates the noise characteristics of the 127-qubit IBM Brisbane device. We applied the same experimental configurations as in the noiseless setting to enable a direct comparison. Among the ten tested conditions involving 4-qubit NQE and QCNN models, five conditions that incorporated either the ZZ feature map or the angle encoding yielded higher average classification accuracy than their classical counterparts. Specifically, the NQE with the ZZ feature map paired with a QCNN including 30 trainable parameters achieved an accuracy of 0.8759, while the same encoding with 8 parameters resulted in 0.8764. The angle encoding conditions also demonstrated notable performance, with 30, 20, and 12 trainable parameters yielding accuracy values of 0.8999, 0.8975, and 0.7015, respectively. In the case of 8-qubit models, seven configurations were examined, among which four conditions exhibited improved performance over the classical neural network model. These included the NQE with ZZ feature map combined with QCNNs containing 45, 30, and 18 trainable parameters, which resulted in classification accuracies of 0.9116, 0.9065, and 0.9101, respectively. Additionally, the NQE with angle encoding and a QCNN with 45 trainable parameters achieved an accuracy of 0.9101. A comprehensive summary of these results comparing 4- and 8-qubit NQE and QCNN models under quantum hardware noise is provided in Figure 15 and Figure 16.

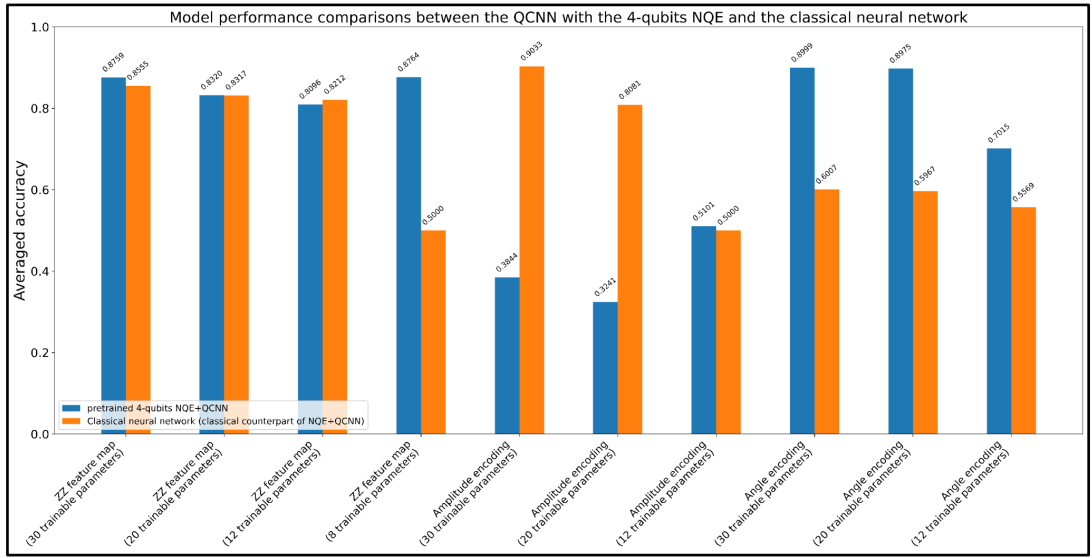

Figure 15. Model performance comparisons between QCNN models with 4-qubits NQE and classical neural network models in the noisy simulation

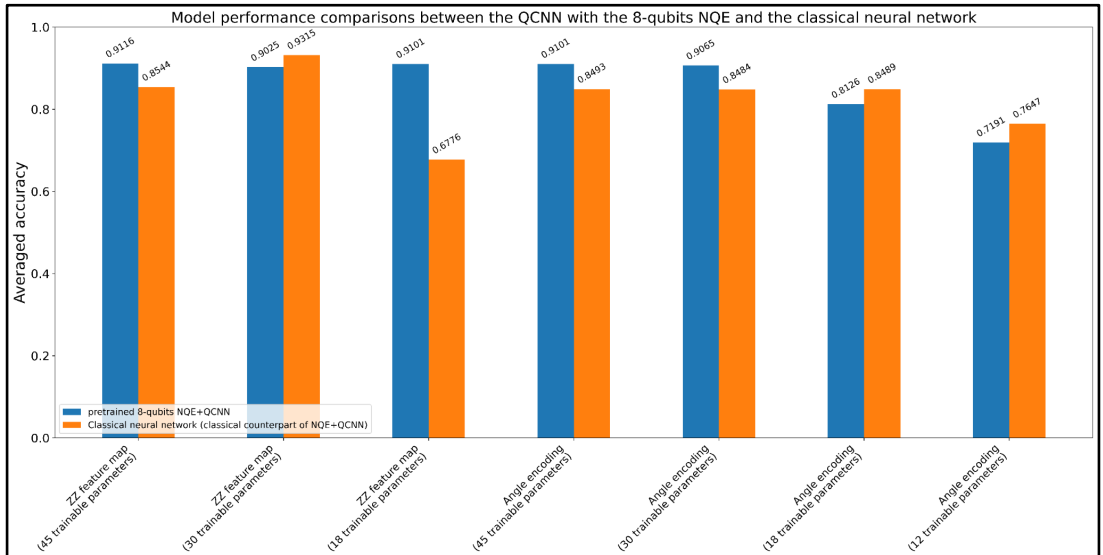

Figure 16. Model performance comparisons between QCNN models with 8-qubits NQE and classical neural network models in the noisy simulation

### 3.4.2 Performance of QCNN with NQE and traditional embeddings including PCA in the noisy environment

In addition to the comparison between classical neural networks and QCNN models with NQE under hardware noise, we further assessed the impact of different quantum embedding methods by comparing the performance of QCNN models using NQE versus those utilizing traditional embeddings. For the 4-qubit QCNN models, we observed that when amplitude encoding was employed, the QCNN models using PCA (i.e., without NQE) outperformed those with NQE in terms of average classification accuracy. In contrast, when using either the ZZ feature map or the angle encoding, the models incorporating NQE consistently demonstrated higher performance compared to the traditional embedding configurations. For the 8-qubit QCNN models, the advantage of NQE was even more pronounced. Across all tested conditions, the models utilizing NQE achieved higher average accuracy than those using traditional embeddings, regardless of the encoding strategy. The full set of comparison results for both 4- and 8-qubit conditions is illustrated in Figure 17 and Figure 18.

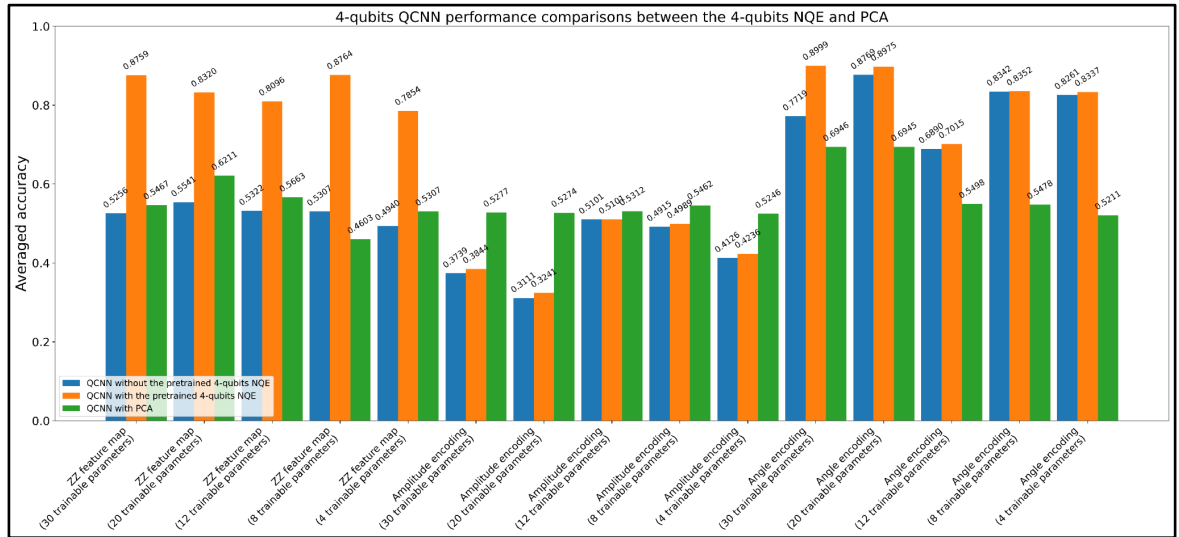

Figure 17. 4-qubits QCNN performances from 4-qubits NQE and PCA conditions in the noisy simulation

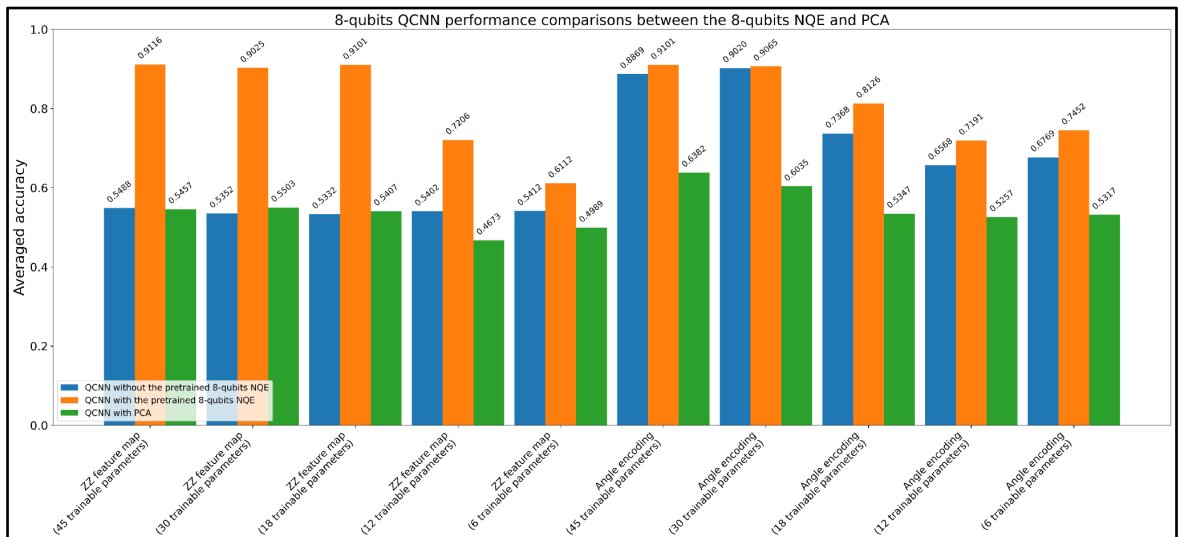

Figure 18. 8-qubits QCNN performances from 8-qubits NQE and PCA conditions in the noisy simulation

*3.4.3 Impact of quantum hardware noises to the QCNN performance trend with the increasing parameter scale*

To investigate the influence of quantum hardware noise on the relationship between model performance and the number of trainable parameters, we compared the performance of QCNN models across different parameter scales (4, 8, 12, 20, and 30 parameters). Unlike the trends observed in the noiseless simulation, no consistent relationship was found between model performance and the increasing number of trainable parameters under noise. In the 4-qubit QCNN configurations with pretrained NQE, despite the presence of noise, models using the ZZ feature map or angle encoding achieved average accuracy values exceeding 0.7. In contrast, the models with amplitude encoding showed the lowest performance, with accuracy values falling below 0.5. A similar pattern was observed in the 8-qubit configurations. The QCNN models with amplitude encoding showed average accuracy values below 0.54, whereas those with ZZ feature map or angle encoding achieved values above 0.6. Notably, in the conditions employing the ZZ feature map and amplitude encoding, the model performance exhibited an increasing trend with the number of trainable parameters. The performance trends across different configurations are illustrated in Figure 19 and Figure 20.

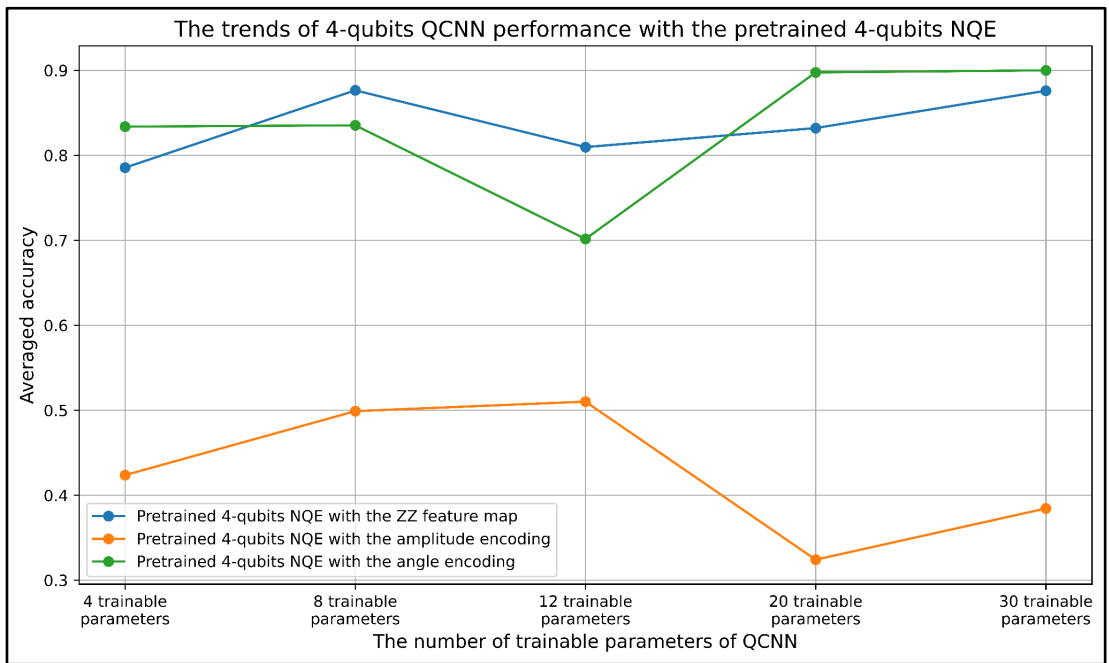

Figure 19. The performance trends of 4-qubits QCNN with the pretrained 4-qubits NQE in the noisy simulation

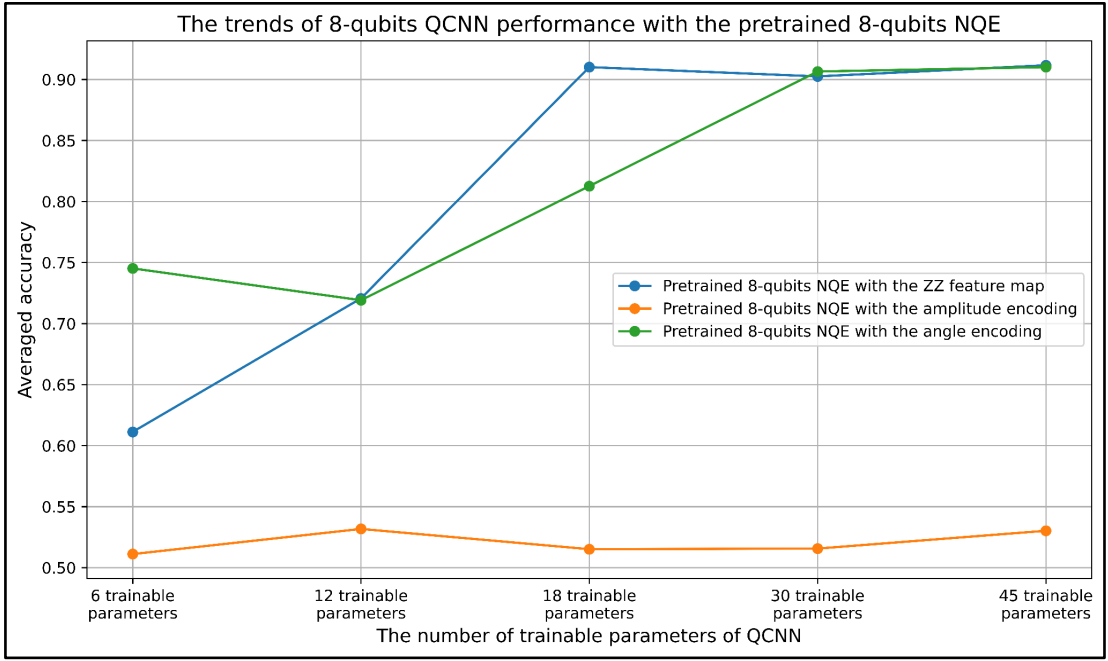

Figure 20. The performance trends of 8-qubits QCNN with the pretrained 8-qubits NQE in the noisy simulation

## 4. Discussions

To propose a QML-based framework for HIV-1 protease cleavage site classification, we adopted the QCNN as the primary quantum model [40, 34]. Using the QCNN algorithm, we aimed to systematically investigate three major aspects that potentially affect model performance: (1) the selection of quantum embedding methods, (2) the influence of trainable parameter scale, and (3) the robustness of QML models under quantum hardware noise.

First, to evaluate the impact of quantum embedding choices, we compared the performance of the NQE method with three traditional alternatives (ZZ feature map, amplitude encoding, and angle encoding). Second, to understand how the number of trainable parameters in the QCNN influences model performance, we implemented five different ansatz configurations with varying parameter scales. Third, we tested all experimental conditions in both noiseless and noisy environments to assess the robustness of each configuration against quantum hardware noise.

In the following subsections, we discuss these three points in detail. We begin by comparing the performance of the QCNN model with its classical neural network counterpart to justify quantum application. We then examine the impact of quantum embedding methods with model accuracy values. Finally, we explore the relationship between the number of trainable parameters and model performance, particularly under realistic noisy conditions.

### 4.1 Classification performances of the QCNN in the noiseless and the noisy simulation

Beyond the simple application of quantum algorithms, it is essential to show a clear rationale for their application. In this regard, previous studies have commonly benchmarked the performance of quantum models against their classical counterparts. For instance, in previous work assessing the feasibility of applying QSVM for core biomarker detection, the authors compared its classification performance directly with that of a classical SVM model [41]. This comparison demonstrated that quantum kernel methods can capture latent patterns that may not be discernible via classical kernels. Similarly, in a study aimed at predicting proton affinity, a hybrid QNN integrating classical neural networks with quantum circuits was introduced [42]. This hybrid architecture achieved comparable performance while requiring significantly fewer parameters than the classical model—2,625 trainable parameters in the classical model versus less than half in the hybrid variant. These results highlighted the potential of quantum circuits to offer computationally efficient alternatives to overparameterized classical networks. A related study proposed a hybrid model for one-dimensional time-series analysis, replacing the classical convolutional layer with a QCNN, and observed improved performance with a reduced number of trainable parameters [22].

Inspired by these findings, we constructed classical neural network models with parameter counts closely matched to those of the QCNN architectures for direct performance comparison. In our experiments, we implemented five different QCNN ansatz, each with a distinct number of parameters: SU4 (15 parameters), U6 (10 parameters), U13 (6 parameters), U15 (4 parameters), and TTN (2 parameters). These were evaluated on both 4- and 8-qubit QCNN models, yielding configurations with 30, 20, 12, 8, and 4 parameters for 4-qubits, and 45, 30, 18, 12, and 6 parameters for 8-qubits, respectively. For the classical counterparts, we designed classical neural networks with comparable parameter counts. Specifically, the 4-qubit QCNN models were matched with classical models having 30, 20, 12, and 8 parameters, while for the 8-qubit QCNNs, classical models with 46, 30, 18, and 12 parameters were used. However, due to constraints in the implementation by the PyTorch framework, some parameter configurations could not be exactly matched or implemented. Accordingly, a few classical counterparts were omitted (please check Tables in Appendix B and C).

In the noiseless simulation environment, QCNN models with NQE using angle and amplitude encoding outperformed their classical counterparts, particularly for the 4-qubit conditions (please check Figure 9). Notably, all QCNN models with angle encoding NQE showed consistently higher average accuracy across all parameter settings. In contrast, the performance of models employing the ZZ feature map or amplitude encoding NQE was either similar to or lower than that of their classical analogs. For the 8-qubit models, QCNNs with ZZ feature map NQE and parameter counts of 45 and 18 achieved higher average accuracy than the corresponding classical models (please check Figure 10). Furthermore, the angle encoding NQE consistently led to improved performance across all 8-qubit QCNN conditions. These results suggest that the proposed QCNN-based framework can deliver higher model performance to classical neural networks under equivalent parameter scales, supporting its potential application for latent pattern analysis in biomedical datasets without relying on hybrid quantum-classical approaches [43, 44].

In the noisy simulation environment, among the 4-qubit conditions, only the QCNN models with angle encoding NQE achieved significantly higher performance than the classical neural networks (please check Figure 15). Moreover, QCNN models with the ZZ feature map and amplitude encoding NQE performed similarly or worse. In the 8-qubit conditions, consistent with the noiseless simulation, the models with ZZ feature map NQE at 45 and 18 parameters QCNN again showed higher averaged accuracy values than the classical models. The same trend was observed for the angle encoding NQE at these parameter settings (please check Figure 16). Taken together, these findings demonstrate that the QCNN architecture retains robustness against quantum hardware noise including T1, T2 decoherence error, and measurement errors. This robustness may stem from the use of only nearest-neighbor two-qubit gates, which helps mitigate SWAP operation errors stemming from limited qubit connectivity [45, 46].

### 4.2 Effectiveness of NQE for the HIV-1 cleavage sites classification with the QCNN

Unlike the traditional embedding methods widely adopted in previous studies, the effectiveness of NQE has been demonstrated in several recent applications. Beyond the original validation of NQE with the QCNN algorithm in the original paper [26], its applicability has also been verified using the deterministic quantum computation with one qubit (DQC1) algorithm on a nuclear magnetic resonance (NMR) quantum processor [47]. In that study, DQC1 with NQE achieved 0.98 classification accuracy on the MNIST dataset, whereas the same protocol without NQE yielded only 0.54 accuracy. These results indicate that NQE enables effective learning even in resource-limited one-qubit quantum algorithms for binary classification tasks. As another example, in a study on gene expression biomarker identification for clear cell renal cell carcinoma (ccRCC) metastasis, a binary classification problem, both QSVM and QNN models using NQE were proposed [48]. The authors reported improved accuracy when NQE was applied: from 0.86 to 0.87 in the QSVM, and from 0.74 to 0.85 in the classical SVM. These findings collectively suggest that NQE contributes to enhanced model performance regardless of whether the model is quantum or classical.

Motivated by these previous results, we compared the performance of QCNN models utilizing pretrained NQE with those using three conventional embedding methods combined with PCA. Before applying the pretrained NQE, we assessed the model optimization of the training through loss value trajectories and trace distance improvements. In the noiseless simulation, all training and validation loss graphs demonstrated stable convergence (please check Figures 5 and 6). Furthermore, the trace distance—initially near zero—consistently increased to values greater than 0.75 after training, for both training and test datasets, with the exception of the 4-qubit ZZ feature map NQE condition (please check Table 4). Based on these observations, we concluded that the NQEs were successfully trained using one-hot encoded amino acid sequences.

In the noisy simulation environment using the IBM FakeBrisbane backend, similar convergence was observed in the training and validation loss graphs. However, in the amplitude encoding NQE condition, gradient explosion occurred, resulting in NaN values during training. Despite this instability, the validation loss continued to decrease over time (please check Figure 7 and Figure 8). While the trace distance improvements were generally lower than those in the noiseless case, they still reached around 0.65. In contrast, the amplitude encoding NQE yielded significantly lower trace distances, around 0.25 or 0.18 after training, indicating a greater vulnerability to hardware-induced noise.

Interestingly, for the 4-qubit ZZ feature map NQE condition, the NQE showed better convergence in the noisy simulation than in the noiseless one. Specifically, the minimum validation loss decreased from 0.25 (noiseless) to below 0.15 (noisy), and the trace distance after training increased dramatically from 0.0854 (training dataset) and 0.0986 (test dataset) in the noiseless case to 0.6658 and 0.5715, respectively, in the noisy case. This trend aligns with earlier reports that certain types of quantum noise may assist optimization in QML algorithms. For instance, previous studies have shown that amplitude damping noise can improve QML performance [49], and that controllable quantum noise may serve a regularization role in QNN training, improving the mean squared error by up to 0.8 [50].

Following the pretraining of the NQE models, we evaluated QCNN performance with the pretrained NQE in comparison to the three conventional embeddings with PCA. In the 4-qubit QCNN models under the noiseless simulation, conditions using NQE with amplitude and angle encoding achieved higher average accuracy than the conventional embeddings. However, the ZZ feature map NQE showed only comparable performance, despite achieving trace distance values around 0.80. Among the 8-qubit QCNN models, overall performance improved when using NQE compared to traditional embeddings. In the noisy simulation, similar trends were observed. The majority of 4- and 8-qubit QCNN models using NQE outperformed those with conventional embeddings, apart from the 4-qubit amplitude encoding NQE condition. These results collectively suggest that NQE enhances QCNN performance in both noiseless and noisy environments, affirming its robustness and adaptability across different experimental settings.

### 4.3 Changes of the QCNN model performance with the increasing trainable parameter scale

In classical machine learning (CML), the relationship between model scale and generalization performance has been studied extensively through large-scale empirical analyses. One such study demonstrated that model performance often follows a power-law scaling with both the model size and the training dataset size, particularly in overparameterized regimes [51]. Similar findings have been reported in the domain of large language models, where performance consistently improves with increased model capacity when sufficient data are available [52]. Furthermore, the phenomenon of double descent has been observed, wherein model performance initially degrades near the interpolation threshold but improves again as the model size increases beyond that point [53].

Motivated by these findings, we investigated whether similar trends hold for QCNN models by examining performance across conditions with varying numbers of trainable parameters. In the noiseless simulation, particularly under the 4-qubit ZZ feature map NQE condition, we observed that increasing the parameter count (from 4 to 30) led to a corresponding increase in average accuracy, even though the trace distance improvement remained minimal. On the other hand, for the amplitude encoding and angle encoding NQE conditions, the performance remained stable across models with different parameter sizes, consistently yielding average accuracy values above 0.85. A similar pattern emerged in the 8-qubit QCNN experiments, where all NQE conditions, except those with only 4 trainable parameters, produced accuracy values exceeding 0.85 (please check Figures 13 and 14). Given that a classical linear SVM baseline achieved an AUC value greater than 0.96 [54], we speculate that these high accuracy scores from QCNN models may reflect the relatively simple structure of the analyzed datasets.

Under noisy simulation with quantum hardware noise, however, averaged accuracy value trends became more fluctuated (please check Figure 19 and Figure 20). In the amplitude encoding NQE condition, both 4- and 8-qubit QCNN models exhibited a notable performance drop, with average accuracy values falling below 0.55. In contrast, all 4-qubit QCNN models using the ZZ feature map and angle encoding NQE maintained accuracy values above 0.70, demonstrating a certain level of resilience to quantum noise. For 8-qubit QCNNs using the ZZ feature map and angle encoding NQE, we observed clear upward trends in performance as the number of trainable parameters increased. These findings collectively suggest that the scalability of QCNN performance with model size may also extend to noisy environments, and that the algorithm maintains a degree of robustness against typical quantum hardware noise, such as decoherence and measurement errors [55].

## 5. Conclusion

In this study, we proposed a QCNN-based classification framework for identifying HIV-1 protease cleavage sites. To enhance model performance, we incorporated the NQE method and systematically compared its effectiveness against conventional quantum embedding techniques, including the ZZ feature map, amplitude encoding, and angle encoding. Furthermore, we investigated the effect of trainable parameter scale on QCNN performance by evaluating five different ansatz configurations.

Our results highlight three key strengths of the proposed framework. First, to the best of our knowledge, this is the first study to apply and evaluate the QCNN algorithm for HIV-1 protease cleavage site prediction under both noiseless and noisy quantum simulation environments. Despite the presence of quantum hardware noise, the QCNN models maintained robust performance, demonstrating their feasibility for biomedical data analysis. Second, we validated the applicability of NQE to biomedical data beyond previously studied toy datasets such as MNIST. Our results suggest that NQE is a viable and effective embedding method for researchers aiming to apply quantum models to biomedical classification tasks. Third, by comparing QCNN models with varying parameter scales, we explored the relationship between model complexity and performance. This analysis provides empirical support for theoretical expectations about model scaling trends in quantum machine learning. Nonetheless, the framework has not yet been externally validated using new sequences or through downstream biological experiments. The QCNN models were trained and evaluated solely on amino acid sequences collected from previous reports. Although these sequences include both viral and host proteins to mitigate source bias, external validation remains necessary.

## Funding
This research was supported by the education and training program of the Quantum Information Research Support Center, funded through the National research foundation of Korea (NRF) by the Ministry of science and ICT (MSIT) of the Korean government (No. RS-20230NR057243) to JC and JL.

## Code availability
The implementation of the NQE can be found at https://github.com/qDNA-yonsei/Neural-Quantum-Embedding. The datasets for the HIV-1 protease cleavage site detection can be downloaded at https://archive.ics.uci.edu/dataset/330/hiv+1+protease+cleavage.

## Conflict of interests
None.

## Author contribution
J.C. designed the study and performed the analysis. J.L. and K.L.J. assisted with interpretation and revision. J.U.J. supervised the project. J.C. drafted the manuscript, and all authors revised and approved the final version.

# Appendix A. The 4-qubits and 8-qubits QCNN model performance with/without NQE

Table A.1 4-qubits QCNN performances with/without pretrained NQE model in the noiseless simulation

| No. | NQE condition | QCNN condition (ansatz / # of trainable parameters) | Hyperparameters for QCNN training (batch size / learning rate / iterations) | Accuracy (mean ± standard deviation) |
|---|---|---|---|---|
| 1 | 4-qubits ZZ[1] | SU4 gate / 30 trainable parameters | 25 / 0.01 / 2000 | 0.5256 ± 0.0147 (without pretraining) <br> 0.6201 ± 0.0289 (with pretraining) |
| 2 | | U6 gate / 20 trainable parameters | 25 / 0.007 / 2000 | 0.5321 ± 0.0080 (without pretraining) <br> 0.6075 ± 0.0336 (with pretraining) |
| 3 | | U13 gate / 12 trainable parameters | 25 / 0.009 / 2000 | 0.5226 ± 0.0010 (without pretraining) <br> 0.5307 ± 0.0289 (with pretraining) |
| 4 | | U15 gate / 8 trainable parameters | 25 / 0.01 / 2000 | 0.5055 ± 0.0072 (without pretraining) <br> 0.5246 ± 0.0040 (with pretraining) |
| 5 | | TTN gate / 4 trainable parameters | 25 / 0.02 / 2000 | 0.5055 ± 0.0033 (without pretraining) <br> 0.5201 ± 0.0100 (with pretraining) |
| 6 | 4-qubits Amp[2] | SU4 gate / 30 trainable parameters | 32 / 0.01 / 2000 | 0.8297 ± 0.0012 (without pretraining) <br> 0.8859 ± 0.0013 (with pretraining) |
| 7 | | U6 gate / 20 trainable parameters | 25 / 0.01 / 2000 | 0.8085 ± 0.1542 (without pretraining) <br> 0.8849 ± 0.0010 (with pretraining) |
| 8 | | U13 gate / 12 trainable parameters | 16 / 0.005 / 2000 | 0.8075 ± 0.1537 (without pretraining) <br> 0.8844 ± 0.0015 (with pretraining) |
| 9 | | U15 gate / 8 trainable parameters | 25 / 0.02 / 2000 | 0.8094 ± 0.1547 (without pretraining) <br> 0.8859 ± 0.0101 (with pretraining) |
| 10 | | TTN gate / 4 trainable parameters | 25 / 0.005 / 2000 | 0.8095 ± 0.1517 (without pretraining) <br> 0.8869 ± 0.0101 (with pretraining) |
| 11 | 4-qubits Ang[3] | SU4 gate / 30 trainable parameters | 25 / 0.006 / 2000 | 0.8392 ± 0.1696 (without pretraining) <br> 0.9246 ± 0.1102 (with pretraining) |
| 12 | | U6 gate / 20 trainable parameters | 25 / 0.01 / 2000 | 0.7528 ± 0.3386 (without pretraining) <br> 0.9226 ± 0.0010 (with pretraining) |
| 13 | | U13 gate / 12 trainable parameters | 25 / 0.03 / 2000 | 0.7568 ± 0.3356 (without pretraining) <br> 0.9246 ± 0.0011 (with pretraining) |
| 14 | | U15 gate / 8 trainable parameters | 25 / 0.01 / 2000 | 0.9261 ± 0.0020 (without pretraining) <br> 0.9261 ± 0.0012 (with pretraining) |
| 15 | | TTN gate / 4 trainable parameters | 25 / 0.005 / 2000 | 0.9231 ± 0.0037 (without pretraining) <br> 0.9251 ± 0.0019 (with pretraining) |

[1] 4-qubits ZZ: NQE with the 4-qubits ZZ feature map; [2] 4-qubits Amp: NQE with the 4-qubits amplitude encoding; [3] 4-qubits Ang: NQE with the 4-qubits angle encoding;

Table A.2 8-qubits QCNN performances with/without pretrained NQE model in the noiseless simulation

| No. | NQE condition | QCNN condition (ansatz / # of trainable parameters) | Hyperparameters for QCNN training (batch size / learning rate / iterations) | Accuracy (mean ± standard deviation) |
|---|---|---|---|---|
| 1 | 8-qubits ZZ[1] | SU4 gate / 45 trainable parameters | 32 / 0.01 / 2000 | 0.5512 ± 0.0080 (without pretraining) <br> 0.9005 ± 0.0056 (with pretraining) |
| 2 | | U6 gate / 30 trainable parameters | 25 / 0.005 / 2000 | 0.5342 ± 0.0020 (without pretraining) <br> 0.9045 ± 0.0050 (with pretraining) |
| 3 | | U13 gate / 18 trainable parameters | 25 / 0.007 / 2000 | 0.5352 ± 0.0010 (without pretraining) <br> 0.9030 ± 0.0020 (with pretraining) |
| 4 | | U15 gate / 12 trainable parameters | 25 / 0.01 / 2000 | 0.5507 ± 0.0166 (without pretraining) <br> 0.8945 ± 0.0165 (with pretraining) |
| 5 | | TTN gate / 6 trainable parameters | 16 / 0.008 / 2000 | 0.5312 ± 0.0282 (without pretraining) <br> 0.6287 ± 0.1426 (with pretraining) |
| 6 | 8-qubits Amp[2] | SU4 gate / 45 trainable parameters | 25 / 0.01 / 2000 | 0.8824 ± 0.0253 (without pretraining) <br> 0.8935 ± 0.0020 (with pretraining) |
| 7 | | U6 gate / 30 trainable parameters | 25 / 0.006 / 2000 | 0.8181 ± 0.1590 (without pretraining) <br> 0.8970 ± 0.010 (with pretraining) |
| 8 | | U13 gate / 18 trainable parameters | 32 / 0.01 / 2000 | 0.8176 ± 0.1588 (without pretraining) <br> 0.8960 ± 0.0020 (with pretraining) |
| 9 | | U15 gate / 12 trainable parameters | 16 / 0.008 / 2000 | 0.8146 ± 0.1586 (without pretraining) <br> 0.8945 ± 0.0066 (with pretraining) |
| 10 | | TTN gate / 6 trainable parameters | 25 / 0.01 / 2000 | 0.8136 ± 0.1744 (without pretraining) <br> 0.8920 ± 0.0127 (with pretraining) |
| 11 | 8-qubits Ang[3] | SU4 gate / 45 trainable parameters | 16 / 0.008 / 2000 | 0.8437 ± 0.1015 (without pretraining) <br> 0.8940 ± 0.0040 (with pretraining) |
| 12 | | U6 gate / 30 trainable parameters | 32 / 0.01 / 2000 | 0.7372 ± 0.3121 (without pretraining) <br> 0.8945 ± 0.1100 (with pretraining) |
| 13 | | U13 gate / 18 trainable parameters | 25 / 0.01 / 2000 | 0.7367 ± 0.3131 (without pretraining) <br> 0.8915 ± 0.0029 (with pretraining) |
| 14 | | U15 gate / 12 trainable parameters | 25 / 0.01 / 2000 | 0.7447 ± 0.2932 (without pretraining) <br> 0.8915 ± 0.0010 (with pretraining) |
| 15 | | TTN gate / 6 trainable parameters | 25 / 0.009 / 2000 | 0.7849 ± 0.1094 (without pretraining) <br> 0.7990 ± 0.0452 (with pretraining) |

[1]8-qubits ZZ: NQE with the 8-qubits ZZ feature map; [2]8-qubits Amp: NQE with the 8-qubits amplitude encoding; [3]8-qubits Ang: NQE with the 8-qubits angle encoding;

Table A.3 4-qubits QCNN performances with/without pretrained NQE model in the noisy simulation

| No. | NQE condition | QCNN condition (ansatz / # of trainable parameters) | Hyperparameters for QCNN training (batch size / learning rate / iterations) | Accuracy (mean ± standard deviation) |
|---|---|---|---|---|
| 1 | 4-qubits ZZ[1] | SU4 gate / 30 trainable parameters | 25 / 0.01 / 500 | 0.5267 ± 0.0156 (without pretraining) <br> 0.8759 ± 0.0034 (with pretraining) |
| 2 | | U6 gate / 20 trainable parameters | 10 / 0.01 / 500 | 0.5541 ± 0.1040 (without pretraining) <br> 0.8320 ± 0.0112 (with pretraining) |
| 3 | | U13 gate / 12 trainable parameters | 15 / 0.008 / 500 | 0.5332 ± 0.0080 (without pretraining) <br> 0.8096 ± 0.0249 (with pretraining) |
| 4 | | U15 gate / 8 trainable parameters | 20 / 0.01 / 500 | 0.5307 ± 0.0096 (without pretraining) <br> 0.8764 ± 0.0010 (with pretraining) |
| 5 | | TTN gate / 4 trainable parameters | 25 / 0.01 / 500 | 0.4940 ± 0.0247 (without pretraining) <br> 0.7854 ± 0.0641 (with pretraining) |
| 6 | 4-qubits Amp[2] | SU4 gate / 30 trainable parameters | 10 / 0.015 / 500 | 0.3739 ± 0.0309 (without pretraining) <br> 0.3844 ± 0.0325 (with pretraining) |
| 7 | | U6 gate / 20 trainable parameters | 15 / 0.0095 / 500 | 0.3111 ± 0.0185 (without pretraining) <br> 0.3241 ± 0.0183 (with pretraining) |
| 8 | | U13 gate / 12 trainable parameters | 20 / 0.002 / 500 | 0.5101 ± 0.0111 (without pretraining) <br> 0.5101 ± 0.0023 (with pretraining) |
| 9 | | U15 gate / 8 trainable parameters | 10 / 0.01 / 500 | 0.4915 ± 0.0085 (without pretraining) <br> 0.4989 ± 0.0085 (with pretraining) |
| 10 | | TTN gate / 4 trainable parameters | 13 / 0.003 / 500 | 0.4126 ± 0.0127 (without pretraining) <br> 0.4236 ± 0.0026 (with pretraining) |
| 11 | 4-qubits Ang[3] | SU4 gate / 30 trainable parameters | 25 / 0.01 / 500 | 0.7719 ± 0.2490 (without pretraining) <br> 0.8999 ± 0.0029 (with pretraining) |
| 12 | | U6 gate / 20 trainable parameters | 20 / 0.004 / 500 | 0.8769 ± 0.0332 (without pretraining) <br> 0.8975 ± 0.0033 (with pretraining) |
| 13 | | U13 gate / 12 trainable parameters | 10 / 0.05 / 500 | 0.6890 ± 0.1075 (without pretraining) <br> 0.7015 ± 0.1158 (with pretraining) |
| 14 | | U15 gate / 8 trainable parameters | 13 / 0.005 / 500 | 0.8342 ± 0.0392 (without pretraining) <br> 0.8352 ± 0.0374 (with pretraining) |
| 15 | | TTN gate / 4 trainable parameters | 25 / 0.001 / 500 | 0.8261 ± 0.0868 (without pretraining) <br> 0.8334 ± 0.0029 (with pretraining) |

[1]4-qubits ZZ: NQE with the 4-qubits ZZ feature map; [2]4-qubits Amp: NQE with the 4-qubits amplitude encoding; [3]4-qubits Ang: NQE with the 4-qubits angle encoding;

Table A.4 8-qubits QCNN performances with/without pretrained NQE model in the noisy simulation

| No. | NQE condition | QCNN condition (ansatz / # of trainable parameters) | Hyperparameters for QCNN training (batch size / learning rate / iterations) | Accuracy (mean ± standard deviation) |
|---|---|---|---|---|
| 1 | 8-qubits ZZ[1] | SU4 gate / 45 trainable parameters | 12 / 0.009 / 300 | 0.5488 ± 0.0307 (without pretraining)<br>0.9116 ± 0.0089 (with pretraining) |
| 2 | | U6 gate / 30 trainable parameters | 20 / 0.003 / 300 | 0.5352 ± 0.0370 (without pretraining)<br>0.9025 ± 0.0137 (with pretraining) |
| 3 | | U13 gate / 18 trainable parameters | 25 / 0.005 / 300 | 0.5332 ± 0.0040 (without pretraining)<br>0.9101 ± 0.0068 (with pretraining) |
| 4 | | U15 gate / 12 trainable parameters | 20 / 0.01 / 300 | 0.5402 ± 0.0323 (without pretraining)<br>0.7206 ± 0.1588 (with pretraining) |
| 5 | | TTN gate / 6 trainable parameters | 12 / 0.007 / 300 | 0.5412 ± 0.0390 (without pretraining)<br>0.6116 ± 0.1488 (with pretraining) |
| 6 | 8-qubits Amp[2] | SU4 gate / 45 trainable parameters | 10 / 0.01 / 300 | 0.5236 ± 0.0159 (without pretraining)<br>0.5302 ± 0.0149 (with pretraining) |
| 7 | | U6 gate / 30 trainable parameters | 30 / 0.004 / 300 | 0.5101 ± 0.0376 (without pretraining)<br>0.5156 ± 0.0204 (with pretraining) |
| 8 | | U13 gate / 18 trainable parameters | 13 / 0.004 / 300 | 0.5035 ± 0.0057 (without pretraining)<br>0.5151 ± 0.0257 (with pretraining) |
| 9 | | U15 gate / 12 trainable parameters | 15 / 0.005 / 300 | 0.5121 ± 0.0285 (without pretraining)<br>0.5317 ± 0.0153 (with pretraining) |
| 10 | | TTN gate / 6 trainable parameters | 12 / 0.008 / 300 | 0.5070 ± 0.0169 (without pretraining)<br>0.5111 ± 0.0140 (with pretraining) |
| 11 | 8-qubits Ang[3] | SU4 gate / 45 trainable parameters | 15 / 0.009 / 300 | 0.8869 ± 0.0225 (without pretraining)<br>0.9101 ± 0.0051 (with pretraining) |
| 12 | | U6 gate / 30 trainable parameters | 20 / 0.015 / 300 | 0.9020 ± 0.0089 (without pretraining)<br>0.9065 ± 0.0080 (with pretraining) |
| 13 | | U13 gate / 18 trainable parameters | 13 / 0.02 / 300 | 0.7367 ± 0.1894 (without pretraining)<br>0.8126 ± 0.1569 (with pretraining) |
| 14 | | U15 gate / 12 trainable parameters | 15 / 0.007 / 300 | 0.6568 ± 0.0919 (without pretraining)<br>0.7191 ± 0.2068 (with pretraining) |
| 15 | | TTN gate / 6 trainable parameters | 10 / 0.008 / 300 | 0.6769 ± 0.0932 (without pretraining)<br>0.7452 ± 0.1285 (with pretraining) |

[1]8-qubits ZZ: NQE with the 8-qubits ZZ feature map; [2]8-qubits Amp: NQE with the 8-qubits amplitude encoding; [3]8-qubits Ang: NQE with the 8-qubits angle encoding;

## Appendix B. The model performance of the classical neural network as a classical counterpart of the QCNN with NQE condition

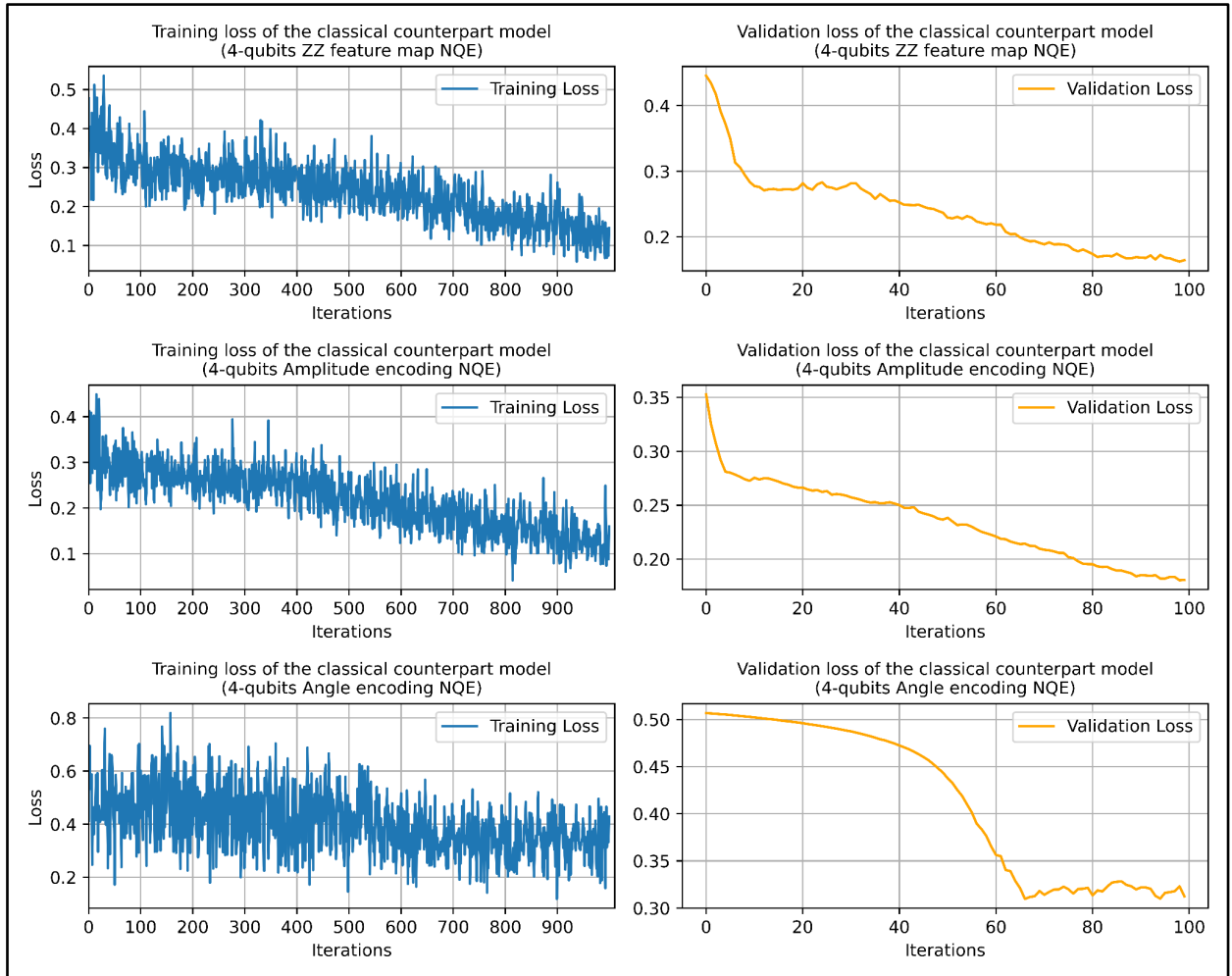

Figure B.1 Training and validation loss graphs of the classical neural network as a classical counterpart of 4-qubits NQE

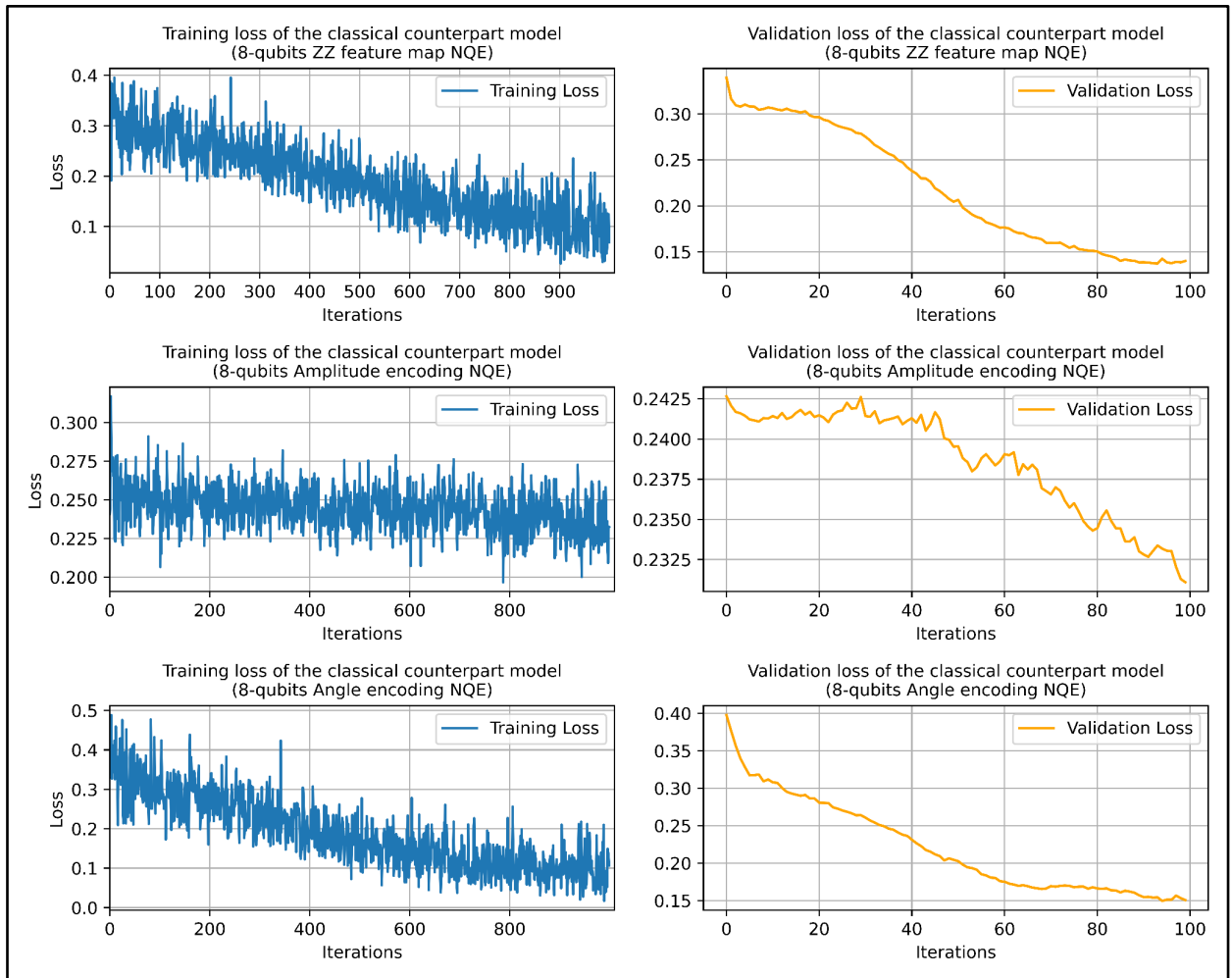

Figure B.2 Training and validation loss graphs of the classical neural network as a classical counterpart of 8-qubits NQE

Table B.1 Classical neural network structures and hyperparameters for the classical counterpart of NQE

| No. | NQE condition | The classical neural network structure | Hyperparameters for the model training |
|---|---|---|---|
| 1 | 4-qubits ZZ feature map | nn.Linear(160, 16),<br>nn.ReLU(),<br>nn.Linear(16, 8) | - batch size: 25<br>- Iterations: 1000<br>- learning rate: 0.005 |
| 2 | 4-qubits amplitude encoding | nn.Linear(160, 32),<br>nn.ReLU(),<br>nn.Linear(32, 16) | - batch size: 25<br>- Iterations: 1000<br>- learning rate: 0.001 |
| 3 | 4-qubits angle encoding | nn.Linear(160, 80),<br>nn.ReLU(),<br>nn.Linear(80, 40),<br>nn.ReLU(),<br>nn.Linear(40, 4) | - batch size: 16<br>- Iterations: 1000<br>- learning rate: 0.0002 |
| 4 | 8-qubits ZZ feature map | nn.Linear(160, 32),<br>nn.ReLU(),<br>nn.Linear(32, 16) | - batch size: 25<br>- Iterations: 1000<br>- learning rate: 0.005 |
| 5 | 8-qubits amplitude encoding | nn.Linear(160, 320),<br>nn.ReLU(),<br>nn.Linear(320, 256) | - batch size: 25<br>- Iterations: 1000<br>- learning rate: 0.003 |
| 6 | 8-qubits angle encoding | nn.Linear(160, 16),<br>nn.ReLU(),<br>nn.Linear(16, 8) | - batch size: 25<br>- Iterations: 1000<br>- learning rate: 0.001 |

Table B.2 Classical neural network hyperparameters and performances for the classical counterpart of the 4-qubits NQE & QCNN

| No. | 4-qubits NQE & QCNN conditions | The classical neural network structure | Hyperparameters (# of trainable parameters) | Accuracy (mean ± standard deviation) |
|---|---|---|---|---|
| 1 | ZZ feature map / SU4 gate | nn.Linear(8, 3, bias=False), nn.ReLU(), nn.Linear(3, 2, bias=False) | - batch size: 128<br>- Iterations: 50<br>- learning rate: 0.0005<br>(30 trainable parameters) | 0.8555 ± 0.0819 |
| 2 | ZZ feature map / U6 gate | nn.Linear(8, 2, bias=False), nn.ReLU(), nn.Linear(2, 2, bias=False) | - batch size: 128<br>- Iterations: 50<br>- learning rate: 0.005<br>(20 trainable parameters) | 0.8317 ± 0.1658 |
| 3 | ZZ feature map / U13 gate | nn.Linear(8, 1, bias=False), nn.ReLU(), nn.Linear(1, 2, bias=True) | - batch size: 128<br>- Iterations: 50<br>- learning rate: 0.004<br>(12 trainable parameters) | 0.8212 ± 0.1615 |
| 4 | ZZ feature map / U15 gate | nn.Linear(8, 1, bias=False) | - batch size: 128<br>- Iterations: 50<br>- learning rate: 0.003<br>(8 trainable parameters) | 0.5000 ± 0.0011 |
| 5 | ZZ feature map / TTN gate | -[1] | - | - |
| 6 | Amplitude encoding / SU4 gate | nn.Linear(16, 2, bias=False) | - batch size: 128<br>- Iterations: 50<br>- learning rate: 0.0005<br>(32 trainable parameters) | 0.9033 ± 0.0203 |
| 7 | Amplitude encoding / U6 gate | nn.Linear(16, 1, bias=False), nn.ReLU(), nn.Linear(1, 1, bias=True) | - batch size: 128<br>- Iterations: 50<br>- learning rate: 0.001<br>(20 trainable parameters) | 0.8081 ± 0.1562 |
| 8 | Amplitude encoding / U13 gate | nn.Linear(16, 1, bias=False) | - batch size: 128<br>- Iterations: 50<br>- learning rate: 0.0005<br>(16 trainable parameters) | 0.5000 ± 0.0011 |
| 9 | Amplitude encoding / U15 gate | - | - | - |
| 10 | Amplitude encoding / TTN gate | - | - | - |
| 11 | Angle encoding / SU4 gate | nn.Linear(4, 4, bias=True), nn.ReLU(), nn.Linear(4, 2, bias=True) | - batch size: 128<br>- Iterations: 50<br>- learning rate: 0.003<br>(30 trainable parameters) | 0.6007 ± 0.0311 |
| 12 | Angle encoding / U6 gate | nn.Linear(4, 3, bias=False), nn.ReLU(), nn.Linear(3, 2, bias=True) | - batch size: 128<br>- Iterations: 50<br>- learning rate: 0.005<br>(20 trainable parameters) | 0.5967 ± 0.0285 |
| 13 | Angle encoding / U13 gate | nn.Linear(4, 2, bias=False), nn.ReLU(), nn.Linear(2, 2, bias=False) | - batch size: 128<br>- Iterations: 50<br>- learning rate: 0.01<br>(12 trainable parameters) | 0.5569 ± 0.0482 |
| 14 | Angle encoding / U15 gate | - | - | - |
| 15 | Angle encoding / TTN gate | - | - | - |

[1]- : In the case of a blank result, this condition was excluded since difficulties in implementing the model using the 'nn.Linear' function of Pytorch with the fixed input dimension.

Table B.3 Classical neural network hyperparameters and performances for the classical counterpart of the 8-qubits NQE & QCNN

| No. | 4-qubits NQE & QCNN conditions | The classical neural network structure | Hyperparameters (# of trainable parameters) | Accuracy (mean ± standard deviation) |
|---|---|---|---|---|
| 1 | ZZ feature map / SU4 gate | nn.Linear(16, 2, bias=True), nn.ReLU(), nn.Linear(2, 3, bias=False), nn.ReLU(), nn.Linear(3, 2, bias=True) | - batch size: 128<br>- Iterations: 50<br>- learning rate: 0.03<br>(50 trainable parameters) | 0.8544 ± 0.1767 |
| 2 | ZZ feature map / U6 gate | nn.Linear(16, 2, bias=False) | - batch size: 16<br>- Iterations: 50<br>- learning rate: 0.001<br>(32 trainable parameters) | 0.9315 ± 0.0218 |
| 3 | ZZ feature map / U13 gate | nn.Linear(16, 1, bias=False), nn.ReLU(), nn.Linear(1, 2, bias=False) | - batch size: 16<br>- Iterations: 50<br>- learning rate: 0.003<br>(18 trainable parameters) | 0.6776 ± 0.2180 |
| 4 | ZZ feature map / U15 gate | -[1] | - | - |
| 5 | ZZ feature map / TTN gate | - | - | - |
| 6 | Angle encoding / SU4 gate | nn.Linear(8, 4, bias=True), nn.ReLU(), nn.Linear(4, 2, bias=True) | - batch size: 16<br>- Iterations: 50<br>- learning rate: 0.001<br>(46 trainable parameters) | 0.8493 ± 0.1768 |
| 7 | Angle encoding / U6 gate | nn.Linear(8, 3, bias=False), nn.ReLU(), nn.Linear(3, 2, bias=False) | - batch size: 16<br>- Iterations: 50<br>- learning rate: 0.001<br>(30 trainable parameters) | 0.8484 ± 0.1749 |
| 8 | Angle encoding / U13 gate | nn.Linear(8, 2, bias=True) | - batch size: 16<br>- Iterations: 50<br>- learning rate: 0.003<br>(18 trainable parameters) | 0.8489 ± 0.1750 |
| 9 | Angle encoding / U15 gate | nn.Linear(8, 1, bias=False), nn.ReLU(), nn.Linear(1, 2, bias=True) | - batch size: 16<br>- Iterations: 50<br>- learning rate: 0.003<br>(12 trainable parameters) | 0.7647 ± 0.2152 |
| 10 | Angle encoding / TTN gate | - | - | - |

[1]- : In the case of a blank result, this condition was excluded since difficulties in implementing the model using the 'nn.Linear' function of Pytorch with the fixed input dimension.

# Appendix C. The hyperparameters and model performances of 4-qubits and 8-qubits QCNN with PCA

Table C.1 4-qubits QCNN performances with PCA (without NQE) in the noiseless simulation

| No. | Quantum embedding | QCNN condition (ansatz / # of trainable parameters) | Hyperparameters for QCNN training (batch size / learning rate / iterations) | Accuracy (mean ± standard deviation) |
|---|---|---|---|---|
| 1 | ZZ feature map | SU4 gate / 30 trainable parameters | 25 / 0.005 / 2000 | 0.6091 ± 0.0244 |
| 2 | | U6 gate / 20 trainable parameters | 32 / 0.01 / 2000 | 0.5990 ± 0.0292 |
| 3 | | U13 gate / 12 trainable parameters | 16 / 0.008 / 2000 | 0.5663 ± 0.0292 |
| 4 | | U15 gate / 8 trainable parameters | 25 / 0.01 / 2000 | 0.4854 ± 0.0316 |
| 5 | | TTN gate / 4 trainable parameters | 32 / 0.007 / 2000 | 0.5186 ± 0.0106 |
| 6 | Amplitude encoding | SU4 gate / 30 trainable parameters | 25 / 0.01 / 2000 | 0.5824 ± 0.0171 |
| 7 | | U6 gate / 20 trainable parameters | 16 / 0.007 / 2000 | 0.5407 ± 0.0075 |
| 8 | | U13 gate / 12 trainable parameters | 25 / 0.01 / 2000 | 0.5302 ± 0.0164 |
| 9 | | U15 gate / 8 trainable parameters | 16 / 0.009 / 2000 | 0.6096 ± 0.0082 |
| 10 | | TTN gate / 4 trainable parameters | 25 / 0.006 / 2000 | 0.5387 ± 0.0312 |
| 11 | Angle encoding | SU4 gate / 30 trainable parameters | 25 / 0.01 / 2000 | 0.4854 ± 0.02255 |
| 12 | | U6 gate / 20 trainable parameters | 32 / 0.005 / 2000 | 0.5799 ± 0.0131 |
| 13 | | U13 gate / 12 trainable parameters | 64 / 0.01 / 2000 | 0.5895 ± 0.0131 |
| 14 | | U15 gate / 8 trainable parameters | 16 / 0.005 / 2000 | 0.5603 ± 0.0214 |
| 15 | | TTN gate / 4 trainable parameters | 25 / 0.01 / 2000 | 0.4498 ± 0.0131 |

Table C.2 8-qubits QCNN performances with PCA (without NQE) in the noiseless simulation

| No. | Quantum embedding | QCNN condition (ansatz / # of trainable parameters) | Hyperparameters for QCNN training (batch size / learning rate / iterations) | Accuracy (mean ± standard deviation) |
|---|---|---|---|---|
| 1 | ZZ feature map | SU4 gate / 45 trainable parameters | 25 / 0.008 / 2000 | 0.5101 ± 0.0122 |
| 2 | | U6 gate / 30 trainable parameters | 16 / 0.01 / 2000 | 0.5112 ± 0.0212 |
| 3 | | U13 gate / 18 trainable parameters | 25 / 0.005 / 2000 | 0.5102 ± 0.0031 |
| 4 | | U15 gate / 12 trainable parameters | 25 / 0.01 / 2000 | 0.5100 ± 0.0032 |
| 5 | | TTN gate / 6 trainable parameters | 32 / 0.005 / 2000 | 0.5100 ± 0.0302 |
| 6 | Angle encoding | SU4 gate / 45 trainable parameters | 25 / 0.005 / 2000 | 0.4448 ± 0.0294 |
| 7 | | U6 gate / 30 trainable parameters | 16 / 0.01 / 2000 | 0.4499 ± 0.0192 |
| 8 | | U13 gate / 18 trainable parameters | 32 / 0.006 / 2000 | 0.4447 ± 0.011 |
| 9 | | U15 gate / 12 trainable parameters | 25 / 0.01 / 2000 | 0.4787 ± 0.0021 |
| 10 | | TTN gate / 6 trainable parameters | 16 / 0.005 / 2000 | 0.4498 ± 0.0101 |

Table C.3 4-qubits QCNN performances with PCA (without NQE) in the noisy simulation

| No. | Quantum embedding | QCNN condition (ansatz / # of trainable parameters) | Hyperparameters for QCNN training (batch size / learning rate / iterations) | Accuracy (mean ± standard deviation) |
|---|---|---|---|---|
| 1 | ZZ feature map | SU4 gate / 30 trainable parameters | 25 / 0.01 / 500 | 0.5467 ± 0.0469 |
| 2 | | U6 gate / 20 trainable parameters | 20 / 0.006 / 500 | 0.6211 ± 0.0357 |
| 3 | | U13 gate / 12 trainable parameters | 25 / 0.009 / 500 | 0.5663 ± 0.0366 |
| 4 | | U15 gate / 8 trainable parameters | 30 / 0.015 / 500 | 0.4603 ± 0.0292 |
| 5 | | TTN gate / 4 trainable parameters | 20 / 0.01 / 500 | 0.5307 ± 0.0075 |
| 6 | Amplitude encoding | SU4 gate / 30 trainable parameters | 15 / 0.011 / 500 | 0.5277 ± 0.0111 |
| 7 | | U6 gate / 20 trainable parameters | 25 / 0.005 / 500 | 0.5277 ± 0.0112 |
| 8 | | U13 gate / 12 trainable parameters | 20 / 0.015 / 500 | 0.5312 ± 0.0044 |
| 9 | | U15 gate / 8 trainable parameters | 15 / 0.0085 / 500 | 0.5462 ± 0.0361 |
| 10 | | TTN gate / 4 trainable parameters | 25 / 0.01 / 500 | 0.5246 ± 0.0558 |
| 11 | Angle encoding | SU4 gate / 30 trainable parameters | 10 / 0.01 / 500 | 0.6946 ± 0.0235 |
| 12 | | U6 gate / 20 trainable parameters | 23 / 0.006 / 500 | 0.6945 ± 0.0168 |
| 13 | | U13 gate / 12 trainable parameters | 30 / 0.02 / 500 | 0.5498 ± 0.0243 |
| 14 | | U15 gate / 8 trainable parameters | 15 / 0.007 / 500 | 0.5478 ± 0.0272 |
| 15 | | TTN gate / 4 trainable parameters | 10 / 0.009 / 500 | 0.5211 ± 0.0484 |

Table C.4 8-qubits QCNN performances with PCA (without NQE) in the noisy simulation

| No. | Quantum embedding | QCNN condition (ansatz / # of trainable parameters) | Hyperparameters for QCNN training (batch size / learning rate / iterations) | Accuracy (mean ± standard deviation) |
|---|---|---|---|---|
| 1 | ZZ feature map | SU4 gate / 45 trainable parameters | 10 / 0.01 / 300 | 0.5457 ± 0.0358 |
| 2 | | U6 gate / 30 trainable parameters | 21 / 0.02 / 300 | 0.5503 ± 0.0120 |
| 3 | | U13 gate / 18 trainable parameters | 15 / 0.007 / 300 | 0.5407 ± 0.0172 |
| 4 | | U15 gate / 12 trainable parameters | 12 / 0.015 / 300 | 0.4673 ± 0.0115 |
| 5 | | TTN gate / 6 trainable parameters | 15 / 0.01 / 300 | 0.5001 ± 0.0272 |
| 6 | Angle encoding | SU4 gate / 45 trainable parameters | 13 / 0.006 / 300 | 0.6382 ± 0.0572 |
| 7 | | U6 gate / 30 trainable parameters | 15 / 0.009 / 300 | 0.6035 ± 0.0523 |
| 8 | | U13 gate / 18 trainable parameters | 20 / 0.01 / 300 | 0.5347 ± 0.0333 |
| 9 | | U15 gate / 12 trainable parameters | 10 / 0.004 / 300 | 0.6257 ± 0.0513 |
| 10 | | TTN gate / 6 trainable parameters | 10 / 0.01 / 300 | 0.5317 ± 0.0191 |

Table C.5 Classical neural network hyperparameters and performances for the classical counterpart of the 4-qubits QCNN with PCA

| No. | QCNN conditions | The classical neural network structure | Hyperparameters (# of trainable parameters) | Accuracy (mean ± standard deviation) |
|---|---|---|---|---|
| 1 | ZZ feature map / SU4 gate | nn.Linear(8, 3, bias=False), nn.ReLU(), nn.Linear(3, 2, bias=False) | - batch size: 16<br>- Iterations: 50<br>- learning rate: 0.003<br>(30 trainable parameters) | 0.7512 ± 0.0602 |
| 2 | ZZ feature map / U6 gate | nn.Linear(8, 2, bias=False), nn.ReLU(), nn.Linear(2, 2, bias=False) | - batch size: 32<br>- Iterations: 50<br>- learning rate: 0.001<br>(20 trainable parameters) | 0.7437 ± 0.0716 |
| 3 | ZZ feature map / U13 gate | nn.Linear(8, 1, bias=False), nn.ReLU(), nn.Linear(1, 2, bias=True) | - batch size: 16<br>- Iterations: 50<br>- learning rate: 0.005<br>(12 trainable parameters) | 0.7416 ± 0.0518 |
| 4 | ZZ feature map / U15 gate | nn.Linear(8, 1, bias=False) | - batch size: 16<br>- Iterations: 50<br>- learning rate: 0.01<br>(8 trainable parameters) | 0.5000 ± 0.0011 |
| 5 | ZZ feature map / TTN gate | -[1] | - | - |
| 6 | Amplitude encoding / SU4 gate | nn.Linear(16, 2, bias=False) | - batch size: 16<br>- Iterations: 50<br>- learning rate: 0.005<br>(32 trainable parameters) | 0.7472 ± 0.0664 |
| 7 | Amplitude encoding / U6 gate | - | - | - |
| 8 | Amplitude encoding / U13 gate | - | - | - |
| 9 | Amplitude encoding / U15 gate | - | - | - |
| 10 | Amplitude encoding / TTN gate | - | - | - |
| 11 | Angle encoding / SU4 gate | nn.Linear(4, 4, bias=True), nn.ReLU(), nn.Linear(4, 2, bias=True) | - batch size: 16<br>- Iterations: 50<br>- learning rate: 0.003<br>(30 trainable parameters) | 0.7140 ± 0.0426 |
| 12 | Angle encoding / U6 gate | nn.Linear(4, 3, bias=False), nn.ReLU(), nn.Linear(3, 2, bias=True) | - batch size: 16<br>- Iterations: 50<br>- learning rate: 0.001<br>(20 trainable parameters) | 0.7049 ± 0.0316 |
| 13 | Angle encoding / U13 gate | nn.Linear(4, 2, bias=False), nn.ReLU(), nn.Linear(2, 2, bias=False) | - batch size: 16<br>- Iterations: 50<br>- learning rate: 0.005<br>(12 trainable parameters) | 0.6667 ± 0.0289 |
| 14 | Angle encoding / U15 gate | nn.Linear(4, 2, bias=False) | - batch size: 16<br>- Iterations: 50<br>- learning rate: 0.003<br>(8 trainable parameters) | 0.6793 ± 0.0150 |
| 15 | Angle encoding / TTN gate | nn.Linear(4, 1, bias=False) | - batch size: 16<br>- Iterations: 50<br>- learning rate: 0.008<br>(4 trainable parameters) | 0.5000 ± 0.0011 |

Table C.6 Classical neural network hyperparameters and performances for the classical counterpart of the 8-qubits NQE with PCA

| No. | 4-qubits NQE & QCNN conditions | The classical neural network structure | Hyperparameters (# of trainable parameters) | Accuracy (mean ± standard deviation) |
|---|---|---|---|---|
| 1 | ZZ feature map / SU4 gate | nn.Linear(16, 2, bias=True), nn.ReLU(), nn.Linear(2, 2, bias=True), nn.ReLU(), nn.Linear(2, 2, bias=True) | - batch size: 16<br>- Iterations: 50<br>- learning rate: 0.001<br>(46 trainable parameters) | 0.6862 ± 0.1226 |
| 2 | ZZ feature map / U6 gate | nn.Linear(16, 2, bias=False) | - batch size: 16<br>- Iterations: 50<br>- learning rate: 0.003<br>(32 trainable parameters) | 0.7507 ± 0.0775 |
| 3 | ZZ feature map / U13 gate | -[1] | - | - |
| 4 | ZZ feature map / U15 gate | - | - | - |
| 5 | ZZ feature map / TTN gate | - | - | - |
| 11 | Angle encoding / SU4 gate | nn.Linear(8, 4, bias=True), nn.ReLU(), nn.Linear(4, 1, bias=True), nn.ReLU(), nn.Linear(1, 2, bias=True), | - batch size: 16<br>- Iterations: 50<br>- learning rate: 0.005<br>(45 trainable parameters) | 0.6370 ± 0.1259 |
| 12 | Angle encoding / U6 gate | nn.Linear(8, 3, bias=False), nn.ReLU(), nn.Linear(3, 2, bias=False) | - batch size: 16<br>- Iterations: 50<br>- learning rate: 0.003<br>(30 trainable parameters) | 0.7523 ± 0.0758 |
| 13 | Angle encoding / U13 gate | nn.Linear(8, 2, bias=True) | - batch size: 16<br>- Iterations: 50<br>- learning rate: 0.001<br>(18 trainable parameters) | 0.7532 ± 0.0664 |
| 14 | Angle encoding / U15 gate | nn.Linear(8, 1, bias=False), nn.ReLU(), nn.Linear(1, 2, bias=True) | - batch size: 16<br>- Iterations: 50<br>- learning rate: 0.001<br>(12 trainable parameters) | 0.7407 ± 0.0519 |
| 15 | Angle encoding / TTN gate | - | - | - |

[1] - : In the case of a blank result, this condition was excluded since difficulties in implementing the model using the 'nn.Linear' function of Pytorch with the fixed input dimension.